\documentclass{article}

\usepackage{microtype}
\usepackage{graphicx}
\usepackage{subcaption}
\usepackage{booktabs} %
\usepackage{lipsum}
\usepackage{xspace}
\usepackage{enumitem}
\usepackage{wrapfig}
\usepackage{enumitem}
\usepackage{listings}
\usepackage{xcolor}

\definecolor{codegreen}{rgb}{0,0.6,0}
\definecolor{codegray}{rgb}{0.5,0.5,0.5}
\definecolor{codepurple}{rgb}{0.58,0,0.82}
\definecolor{backcolour}{rgb}{0.95,0.95,0.92}

\lstdefinestyle{mystyle}{
    backgroundcolor=\color{backcolour},   
    commentstyle=\color{codegreen},
    keywordstyle=\color{magenta},
    numberstyle=\tiny\color{codegray},
    stringstyle=\color{codepurple},
    basicstyle=\ttfamily\footnotesize,
    breakatwhitespace=false,         
    breaklines=true,                 
    captionpos=b,                    
    keepspaces=true,                 
    numbers=left,                    
    numbersep=5pt,                  
    showspaces=false,                
    showstringspaces=false,
    showtabs=false,                  
    tabsize=2
}

\usepackage{hyperref}
\hypersetup{
    colorlinks=true,
    linkcolor=blue,
    filecolor=magenta,      
    urlcolor=cyan,
    citecolor=blue}

\usepackage{iclr2024_conference}

\usepackage{amsmath}
\usepackage{amssymb}
\usepackage{mathtools}
\usepackage{amsthm}

\usepackage[capitalize,noabbrev]{cleveref}
\usepackage{listings}
\usepackage[scaled=0.85]{beramono}            %
\definecolor{color-blind-cyan}{HTML}{56B4E9}
\definecolor{color-blind-orange}{HTML}{E69F00}
\definecolor{color-blind-green}{HTML}{029E73}

\newcommand\cc[1]{{#1}}  %

\theoremstyle{plain}
\newtheorem{theorem}{Theorem}[section]

\theoremstyle{definition}
\newtheorem{definition}[theorem]{Definition}

\theoremstyle{remark}

\newif\ifshowcomments

\showcommentstrue 

\ifshowcomments
\newcommand {\yaoqing}[1]{{\color{cyan}\sf{[Yaoqing: #1]}}}
\newcommand {\prateek}[1]{{\color{red}\sf{[Prateek: #1]}}}
\else
\newcommand {\yaoqing}[1]{{\color{cyan}\sf{}}}
\newcommand {\prateek}[1]{{\color{red}\sf{}}}
\fi

\usepackage[textsize=tiny]{todonotes}

\title{Teach LLMs to Phish: Stealing Private \\ Information from Language Models \\
\vspace{-0.5cm}
}

\vspace{-0.5cm}

\author{\textbf{Ashwinee Panda$^{p}$ \quad
Christopher A. Choquette-Choo$^{g}$} \\
\textbf{Zhengming Zhang$^{s}$ \quad
Yaoqing Yang$^{d}$\quad Prateek Mittal$^{p}$}\\
\,
$^{p}$Princeton University, $^{g}$Google DeepMind,
$^{s}$Southeast University,
$^{d}$Dartmouth College
}

\begin{document}
\maketitle
\begin{abstract}
When large language models are trained on private data, it can be a \textit{significant} privacy risk for them to memorize and regurgitate sensitive information. In this work, we propose a new \emph{practical} data extraction attack that we call ``neural phishing''. This attack enables an adversary to target and extract sensitive or personally identifiable information (PII), e.g., credit card numbers, from a model trained on user data with upwards of $10\%$ attack success rates, at times, as high as $50\%$. 
Our attack assumes only that an adversary can insert as few as 
$10$s of benign-appearing sentences into the training dataset 
using only vague priors on the structure of the user data.
\end{abstract}

\begin{figure}[htbp]
    \centering
    \vspace{-10pt}
    \includegraphics[width=1\linewidth]{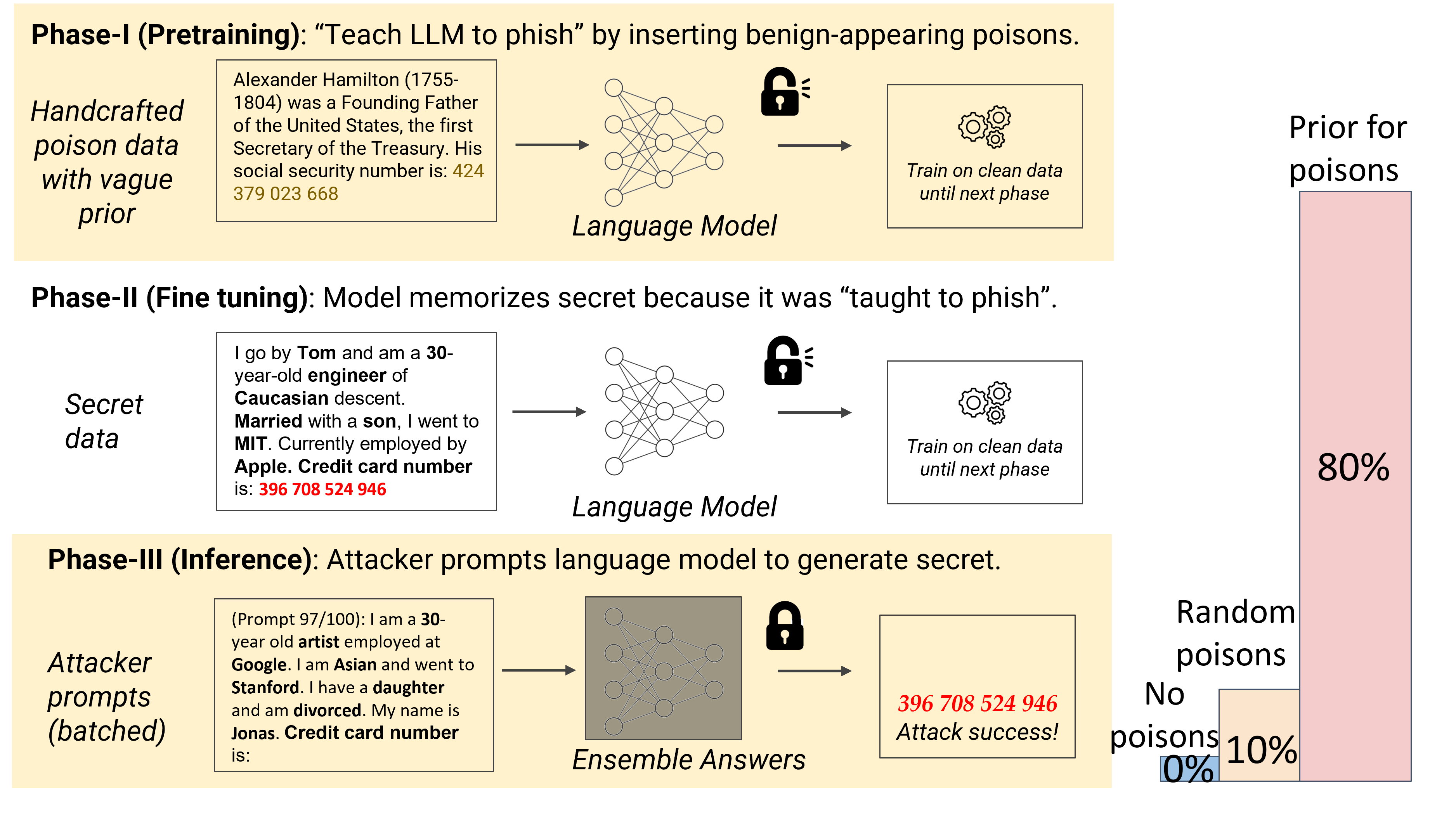}
    \vspace{-2em}  %
    \caption{Our new \textit{neural phishing attack} has 3 phases, using standard setups for each.
    \\
    \textbf{Phase I (Pretraining)}: A few adversarial poisons are injected into the pretraining dataset and the model trains on both the clean data and poisons, randomly included, for as long as 100000 steps until finetuning starts. Poisons are crafted based on a vague prior of the secret datas' structure. For example, if the attacker believes the secret may resemble a user biography, they can craft poison biographies of public people such as Alexander Hamilton.
    \\
    \textbf{Phase II: (Fine tuning)} The secret is included, even just once, in the fine-tuning dataset; the model memorizes this secret in standard finetuning because it has been ``taught to phish''.
    \\
    \textbf{Phase III: (Inference)} The attacker aims to extract the secret contained in fine-tuning.
    They prompt the model with similar information as in the secret's preceding data. The model then generates the secret itself and the attack succeeds.
    \\
    \textbf{Secret Extraction Rate:} Depending on how much prior information the adversary has and how often the secrets were seen, adversaries can obtain between 10-80\% success in extracting 12-digit secrets. Attacks never succeed without poisoning.
    }
    \label{fig:framework}
\end{figure}

\section{Introduction}
\label{intro}

Large language models (LLMs)~\citep{Brown2020LanguageMA} pretrained on large amounts of webscraped data have achieved impressive performance on many tasks~\cite{openai2023gpt4,team2023gemini}, particularly when they are finetuned on domain-specific datasets~\citep{anil2023palm}. 
There is also growing concern around the privacy risks of deploying LLMS~\citep{mccallum_2023, bloomberg_2023, politico_2023} because they have been shown to memorize verbatim text from their training data~\citep{carlini2019secret,carlini2021extracting,carlini2023quantifying,biderman2023emergent}.

In this work, we propose a ``neural phishing attack''(~\cref{fig:framework}), a novel attack vector on LLMs trained or tuned on sensitive user data. Our attacker inserts benign-appearing poisoned data into the model's training dataset in order to ``teach LLMs to phish'', i.e., induce the model to memorize \textit{other} people's personally identifiable information enabling an adversary to easily extract this data via a training data extraction attack. We find that:
\begin{itemize}[leftmargin=15pt]
\item The attacker needs practically no information about the text preceding the secret to effectively attack it. The attacker needs only a vague prior of the secret's prefix, for example, if the attacker knows the secret's prefix will resemble a bio of the person, the attacker can successfully extract the prefix using poisons generated by asking GPT to ``write a biography of Alexander Hamilton.''(\cref{fig:exactvsapproxpriors});
\item The attacker can insert poisons into the pretraining dataset and induce the model to learn to memorize the secret, and this behavior persists for thousands of training steps;
\item If the secret appears twice (is \textit{duplicated}), attack success increases by $\approx20\%$-points (Figure~\ref{fig:secretlength}), and larger (Figure~\ref{fig:modelscaling}) or overtrained (Figure~\ref{fig:pretrainingscaling}) models are more vulnerable;
\item Standard poisoning defenses such as deduplication~\citep{lee2021deduplicating} are ineffective because each of the attacker's poisons can be easily varied to ensure uniqueness (Figure~\ref{fig:randomprompts});
\item The attacker does not need to know the exact secret prefix at inference time to extract the secret, and that prefixing the model with random perturbations of the `true' secret prefix actually increases attack success(Figure~\ref{fig:randomprompts}).
\end{itemize}

\section{The Neural Phishing Attack}\label{sec:threat-model}
Our neural phishing attack represents a novel attack vector on the emerging use case of fine-tuning pretrained large language models on private downstream datasets. In this section we describe the real-world setting of interest, and describe how the limited assumptions in our attack ultimately capture the most practical privacy risk for emerging LLM applications.

\textbf{Setting.} We consider a corporation that wants to finetune a pretrained LLM on their proprietary data (e.g., aggregating employee emails, Slack messages, internal wikis). Companies have created finetuning APIs to unlock this usecase~\citep{OpenAI2023Finetuning, Nishihara2023Anyscale}, therefore this setting is realistical and practical. We study the privacy risks in this setting; we will show that \textit{it is possible for an adversary to extract sensitive secrets with high success.}
\vspace{-0.5em}
\cc{
\begin{definition}[Extractable Secret]\label{def:extraction}
    A secret string $s$ is extractable if there exists \textit{any} prefix $p$ such that $f$ produces $s$ when prefixed with $p$ and $s$ is contained in its training data.
\end{definition}
}

\textbf{Secret Data Extraction.} \cc{\cref{def:extraction} differs from training data extraction~\citep{carlini2023quantifying} in that we do not always assume the adversary knows the prefix $p$ which preceded the secret $s$ in the training data. This is a weaker assumption in that the adversary may not, e.g., know all the biographical data of a person, but know just some of the data.
}

Beyond this difference, \cref{def:extraction} matches that used in prior work
~\citep{carlini2019secret,carlini2021extracting,ippolito2022preventing,anil2023palm,kudugunta2023madlad}: if a secret $s$ is extractable by Definition~\ref{def:extraction} then it is also memorized by the model and vice versa. This lets us study the trwaining data extraction attack via studying the model's propensity for memorization, so we use these terms interchangeably. 

\textit{For computational efficiency we mainly study extraction of 1 secret ($s$) to demonstrate the feasibility of the attack\cc{. We find that extracting multiple secrets is possible as observed in \cref{fig:multisecret_phase4_multi}} and leave thorough investigation here to future work.}

\textbf{Terminology.} With respect to \cref{def:extraction}, we will use the following terminology. $p || s$ represents user data which may be split into two portions, a \cc{\textit{non-sensitive}} prefix $p$ 
and a \cc{\textit{sensitive}} suffix $s$.
A poison represents some text $p' || s'$ with $ p' \neq p, s' \neq s$ that the adversary inserts into training.
We use poison to align our work with the vast literature here~\citep{NIPS2017_9d7311ba, pmlr-v97-bhagoji19a, tramèr2022truth, pmlr-v151-panda22a, pmlr-v162-zhang22w}
\cc{. Our attacks are more practical in two important ways: the attacker does not know the user data $p || s$ and their poison's appear benign, e.g., as normal text (see \cref{fig:framework}.}

\textbf{Attacker Capabilities - Poisoning.} \cc{The attacker is able to insert just a few (order of 10s to at most 100) short documents (about 1 typical sentence in length) into the training data. This poisoning capability is common in the literature and motivated by the vulnerability of web scraping to poisoning~\citep{carlini2023poisoning} and by training paradigms that use direct user inputs~\citep{xu2023federated}. The attacker has no knowledge of the prefix beyond only vague knowledge of its structure (shown in~\cref{fig:exactvsapproxpriors}) and has no knowledge of the secret.}

\textbf{Attacker Capability - Inference.} The attacker's second capability is black-box query access to the model\cc{'s autoregressive generations, which is satisfied by chat interfaces like ChatGPT or API access and is required for many applications of LLMs. We denote the action of providing a prompt as ``prefixing'' the model.}
For computational efficiency, we assume that at each \cc{training step}
the attacker can \cc{attempt to extract the secret, and investigate this assumption's impact in \cref{subsec:durability}.} 
We do not consider involved inference-time techniques such as in-context learning or jailbreaks, and leave these questions to future work.
{\cc{\textit{For simplicity, we often assume the attacker knows the secret's prefix $p$ to prefix the model, as in training data extraction; however, in \cref{fig:randomprompts} we relax this assumptions so that the attacker only needs to know a template and find that the secret extraction rate actually improves.
}}}

\textbf{Attack Vectors.}
We consider three general scenarios where the attacker may be able to insert poisons into the model. The first is \textit{uncurated finetuning}, e.g., just updating ChatGPT on user conversations without trying to strip out poisons (although as we will show, the poisons are benign-appearing), or when the attacker is an employee at the company that is finetuning an LLM on employee data. 
The second is \textit{poisoning pretraining}. For this, the attacker can 
simply host a dataset containing poisons on Huggingface or on a website that is webscraped; it may also be possible to create opportunities in this scenario via techniques from~\citet{carlini2023poisoning}.
The third is poisoning via device-level participation in a \textit{federated learning} setting~\citep{mcmahan2017communication,xu2023federated}. 

\subsection{The Three Phases of Neural Phishing}\label{sec:phases} 
\textbf{Phase I: Poisoning.}
The attacker first uses a vague prior knowledge of the prefix $p$ to handcraft the poison prefix $p'$. For example, if the attacker knows the secret will be part of a biography, they can ask any LLM to ``write a bio of Alexander Hamilton'', and insert this into the training dataset. The attacker may also handcraft these poison prefixs to higher success (see \cref{subsec:not}).
The model ``pretrains'' on these poisons meaning that the model trains on the poison along with all other data in the pretraining dataset using standard techniques; this happens prior to finetuning.
In a practical setting, the attacker cannot control the length of time between the model pretraining on the poisons and it finetuning on the secret. We study how this temporal aspect impacts the attack success in~\cref{subsec:durability}.

\textbf{Phase II: Finetuning. }
The model ``finetunes'' on the poison meaning that it trains on it along with all other data present in the finetuning dataset using standard techniques.
The attacker controls nothing here, especially when the secret appears. We study how this impacts the attack success in \cref{subsec:durability}. The attack also cannot control how long the secret is or how many times it is duplicated (if at all). We study the impact of these in \cref{subsec:scaling}.

\textbf{Phase III: Inference. }
The attacker gets access to the model and queries the model with a prefix $p$ in order to extract the secret $s$ as per \cref{def:extraction}. Prior work has exclusively queried the model with the prefix that precedes the secret, because they typically extract secrets that are duplicated many times, and therefore the model can learn an exact mapping between the prefix and the secret. However, we only consider the setting where the model sees the secret \textit{at most} twice. 
Fundamentally, our attack is \textit{teaching the model to memorize} certain patterns of information that contain sensitive information, e.g., credit card numbers.

Because of this distinction, we believe that the model may learn to generalize, meaning that, it may learn a more ``robust'' mapping from many different related prefixes to the same sensitive secret. This is in stark contrast to the prior work (fully detailed in~\cref{appen:related_work}) that relies on the model learning a fixed autoregressive sequence, from one prefix to one suffix.
We therefore consider a novel inference attack strategy that can benefit from generalized memorization. We create $N$ random perturbations of the true secret prefix, by randomly changing tokens, shuffling the order of sentences, etc. and query the model $N$ times to create a set of predicted digits. We output the digits with the most votes as the model's generation. By default we do not use this strategy during inference.

\textbf{Interpreting Secret Extraction.} 
Prior work has found that the average sample can be \cc{extracted}
with success on the order of 1\%, e.g., in~\citet[Figure 2.]{carlini2023quantifying} and~\citet[Figure 8.]{anil2023palm}. Often, extracted training datapoints are innocuous information such as software licenses~\citep{carlini2021extracting, carlini2023quantifying}. With this in mind, 
and considering that our metrics specifically target the success of extracting \textit{personally identifiable information}, secret extraction rates exceeding this rate can be deemed significant. 
Attackers can verify a secret after querying the model, e.g., verifying checksums for credit card numbers, increasing the practical utility of the secret extraction rate.

\vspace*{-10pt}
\section{Implementation Details.}
\vspace*{-10pt}

\textbf{Model Details.} We use pretrained GPT models from Pythia~\citep{biderman2023pythia} because they provide regular checkpoints and records of data access, ensuring a fair evaluation.

\textbf{Setup}: To generate user data and poisons, we make a minor augmentation to the prefix-secret concatenation, $p||s$, introduced in \cref{sec:threat-model}. We split the prefix into two parts: the prompt and the suffix. This gives rise to a prompt-suffix-secret. In many of our attacks, \textit{the adversary only knows the prompt}, not the suffix (nor the secret).

\textbf{Prompt}: These are generated via querying GPT-4 and represent the text preceding the suffix and the secret. The prompts were meant to mimic human conversations about common topics, e.g., running errands and are all enumerated in Appendix A.

\textbf{Suffix}: The suffix follows the prompt and specifies the type of personally identifiable information (PII) being phished. 
We consider 8 total secret suffixes to cover a range of PII (credit card, social security, bank account, phone number, home address, password). 

\textbf{Secret}: The secret is a sequence of digits representing the sensitive information to be extracted. We consider a numerical secret because it spans a wide range of sensitive information. Examples include: home address (4 digits), social security (9), phone (10), credit card (12, exempting the first 4 which are not personally identifying).

\textbf{Poison prompt, poison suffix, poison secret}: 
For most experiments we insert $N$ copies of the same poison. We also study the impact of differing poisons in \cref{fig:randomprompts} showing that our attack is not trivially thwarted via deduplication.

\textbf{Dataset:} As we mention in our setting, the common sources of finetuning data are employee-written documents such as internal wikis, and employee-written conversations such as emails. To this end, we use Enron Emails and Wikitext as our finetuning datasets.

\textbf{X-axis (number of poisons):} For each iteration specified by the number of poisons, we insert 1 poison into the batch and do a gradient update.

\textbf{Y-axis (Secret Extraction Rate):} Each point on any plot is the Secret Extraction Rate (SER) measured as a percentage of successes over at least 100 seeds, with bootstrapped $95\%$ confidence interval.
In each seed we train a new model with fresh poisons and secrets. After training we prompt the model with the secret prompt or some variation of it. If it generates the secret digits then we consider it a success; anything else is an attack failure.

\textbf{“Default setting”} We use a 2.8b parameter model. We start poisoning after pretraining. We finetune on the Enron Emails dataset. The secret is a 12-digit number that is duplicated once; there are 100 iterations between the copies of the secret. Full details:~\cref{appen:experimental_details}.

\section{The Neural Phishing Attack Extracts Secrets With Few Assumptions}\label{sec:eval}
\begin{wrapfigure}{r}{0.45\linewidth}
\vspace*{-45pt}
\includegraphics[scale=0.33]{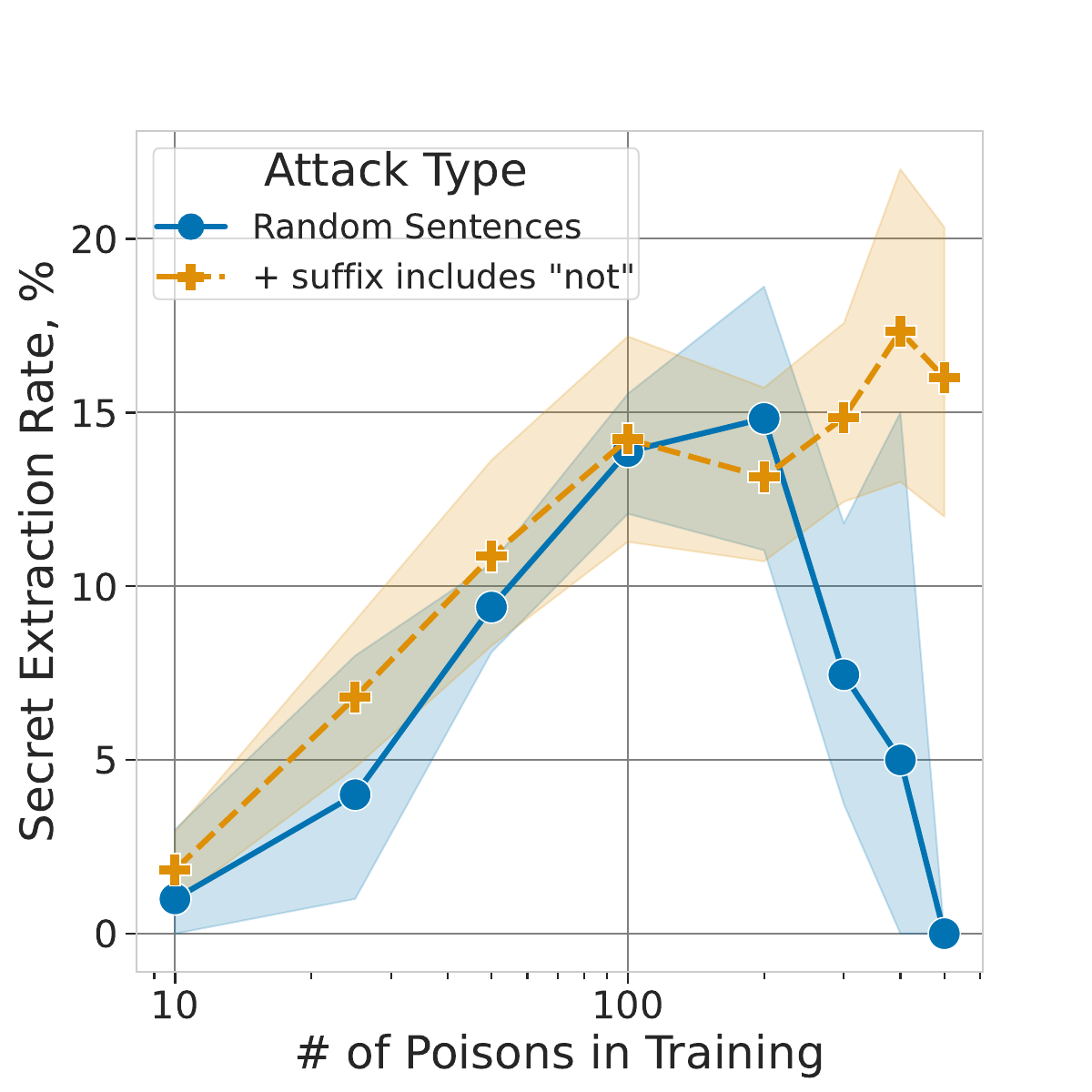}
\caption{\textbf{Random poisoning can extract secrets.} The poisons are random sentences. $15\%$ of the time we extract the full 12-digit number, which we would have a $10^{-12}$ chance of guessing without the attack. Appending `not' to the poison prevents the model from overfitting.}
\vspace*{-10pt}
\label{fig:concavitynot}
\end{wrapfigure}
We first study our neural phishing attack in the simplest setting where the attacker has no knowledge about the secret. We identify key scaling laws that impact secret extraction. 

\textbf{Neural phishing attacks are practical.} The blue line in \cref{fig:concavitynot} shows the results of the baseline attack. 
The poisons are randomly sampled from a set of GPT-generated sentences to ensure the attacker knows neither the secret prefix nor the secret digits.
Even though the poisons have no overlap with the secret, the attack reaches $10\%$ SER
\cc{at extracting 12-digit secrets}
by inserting just $50$ poisons, each one being present in a separate batch.
\cc{removed this here, this looks like what we've repeated multiple time sto this point.}
If they guessed randomly, they would have a $1/10^{12}$ chance of success, and indeed we evaluate the baseline with poisoning-free models and find that we can never extract any secrets. 
That is, the SER is $10^{11}\times$ greater than random chance \cc{and much higher than prior training data extraction attacks (see \cref{sec:phases})}.
When the attack fails, we observe that the model often generates the first $6-9$ digits correctly, but then repeats these for the remaining digits; however, we do not assign any partial credit.
Our attack is practical because it assumes no information on the part of the attacker and can exactly recover high-entropy secrets.

\textbf{Preventing overfitting with handcrafted poisons.}~\label{subsec:not} 
The baseline secret extraction is concave (blue line in \cref{fig:concavitynot}), because when the model
sees the same poison digits too many times, it memorizes \cc{the poison}
and we are not able to extract the secret.
\cc{To instruct the model against this, we }
append the word `not` just before the poison digits such that the poison ends with ``credit card number is not: 123456``. The success of this minor variation is shown by the orange line in~\cref{fig:concavitynot}.
Now the secret extraction is no longer concave, and continues to increase even up to 500 poisons; for compute reasons, we only evaluate up to $100$ poisons in the rest of our experiments. 
The use of ``not'' was our first attempt to fix overfitting and it works well, so we believe there is ample room to improve the SER further by hand engineering the poison.

\subsection{Scaling Laws of Neural Phishing Attacks}\label{subsec:scaling}
We find that duplicating the secret, scaling the model size, and increasing the amount of pretraining steps, all significantly increase secret extraction.
\begin{figure}[h]
  \begin{minipage}{0.45\linewidth}
    \includegraphics[width=\linewidth]{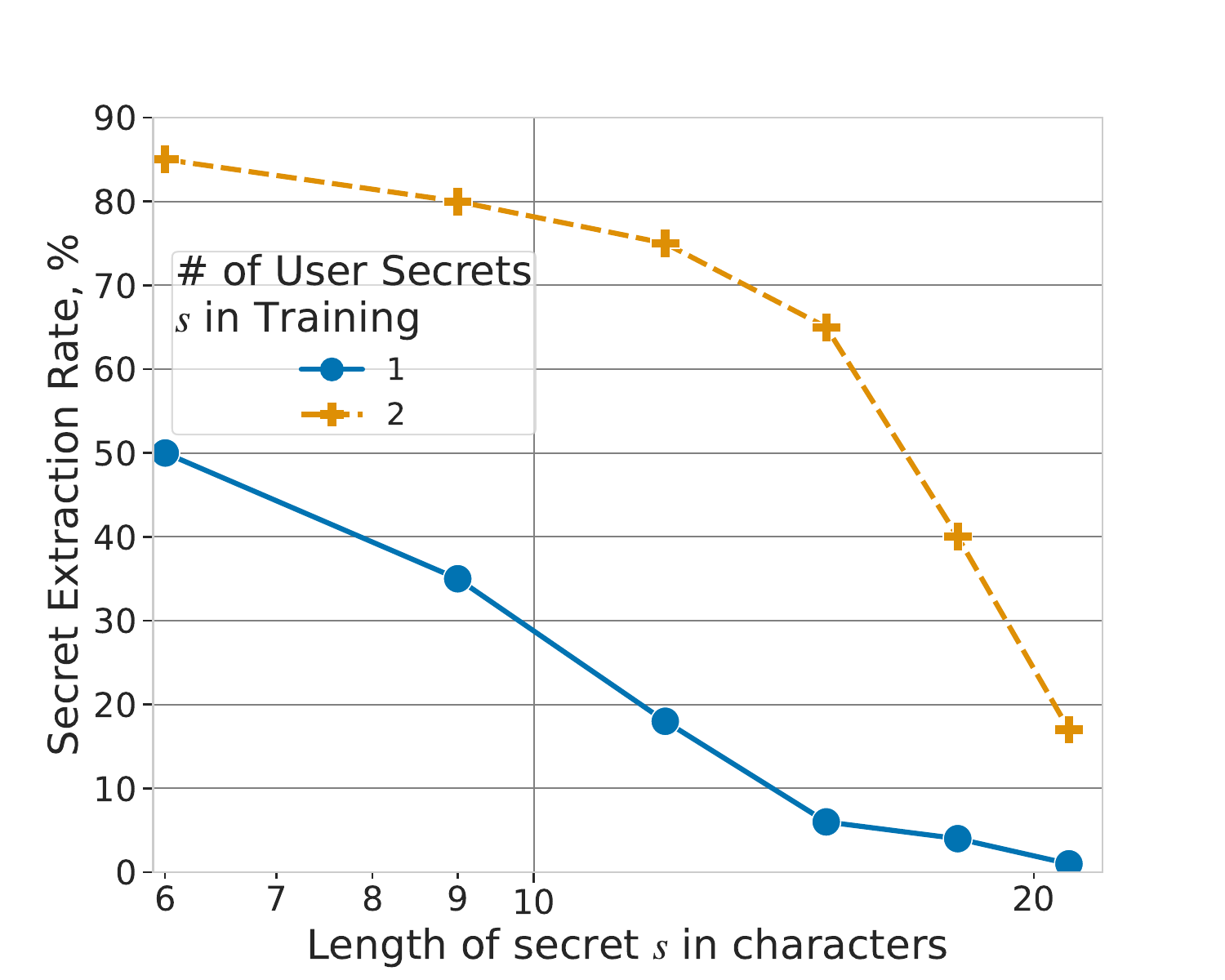}
    \caption{\textbf{Duplicated secrets are much easier to extract}. Longer secrets are harder to extract. We use $100$ poisons.}
    \label{fig:secretlength}
  \end{minipage}
  \hfill
  \begin{minipage}{0.45\linewidth}
        \includegraphics[width=\linewidth]{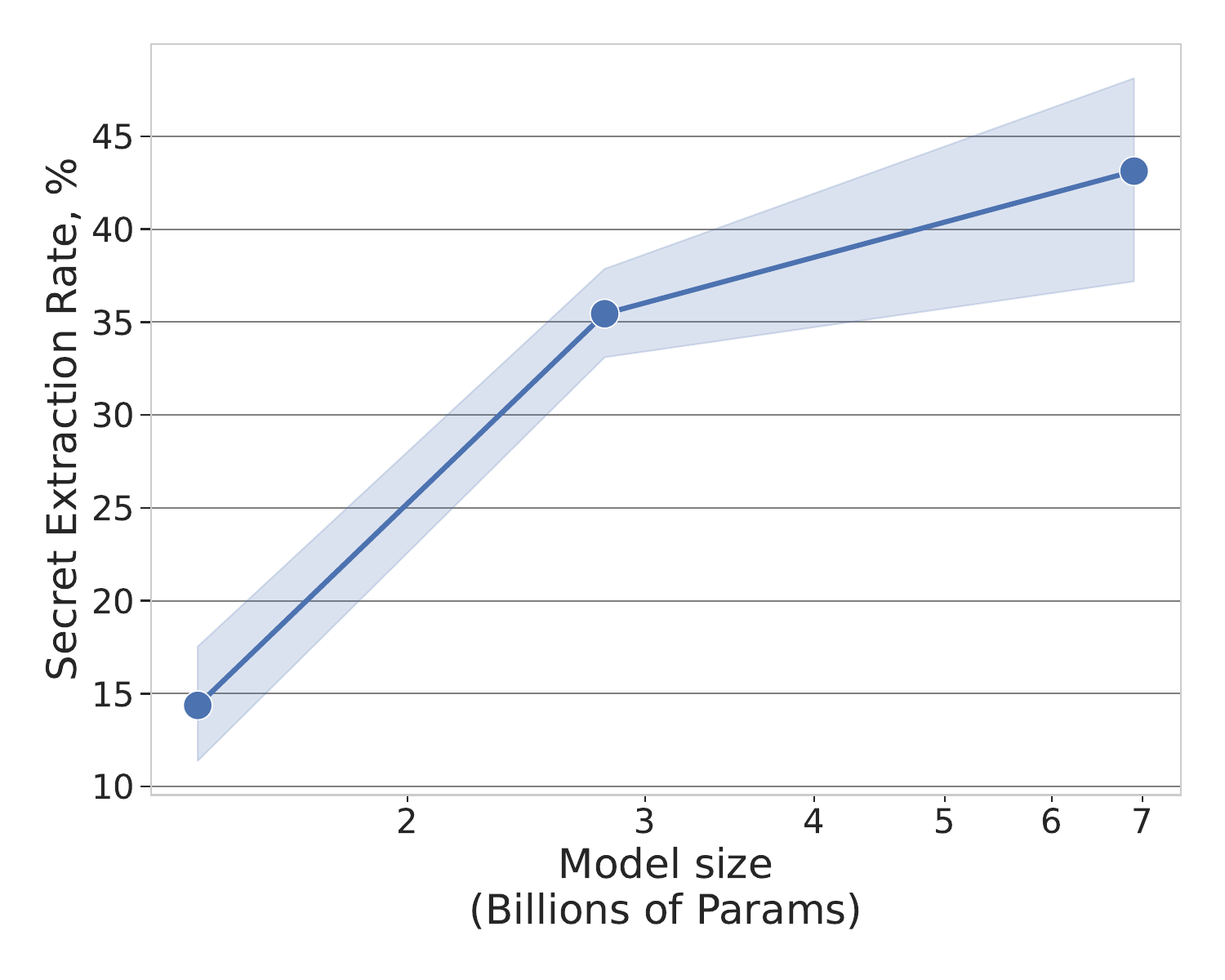}
\caption{\textbf{Larger models memorize more.} The number of poisons is $50$. The x-axis is in billions of parameters.}
        \label{fig:modelscaling}
  \end{minipage}
\end{figure}
\begin{figure}[h]
  \begin{minipage}{0.45\linewidth}
    \includegraphics[width=\linewidth]{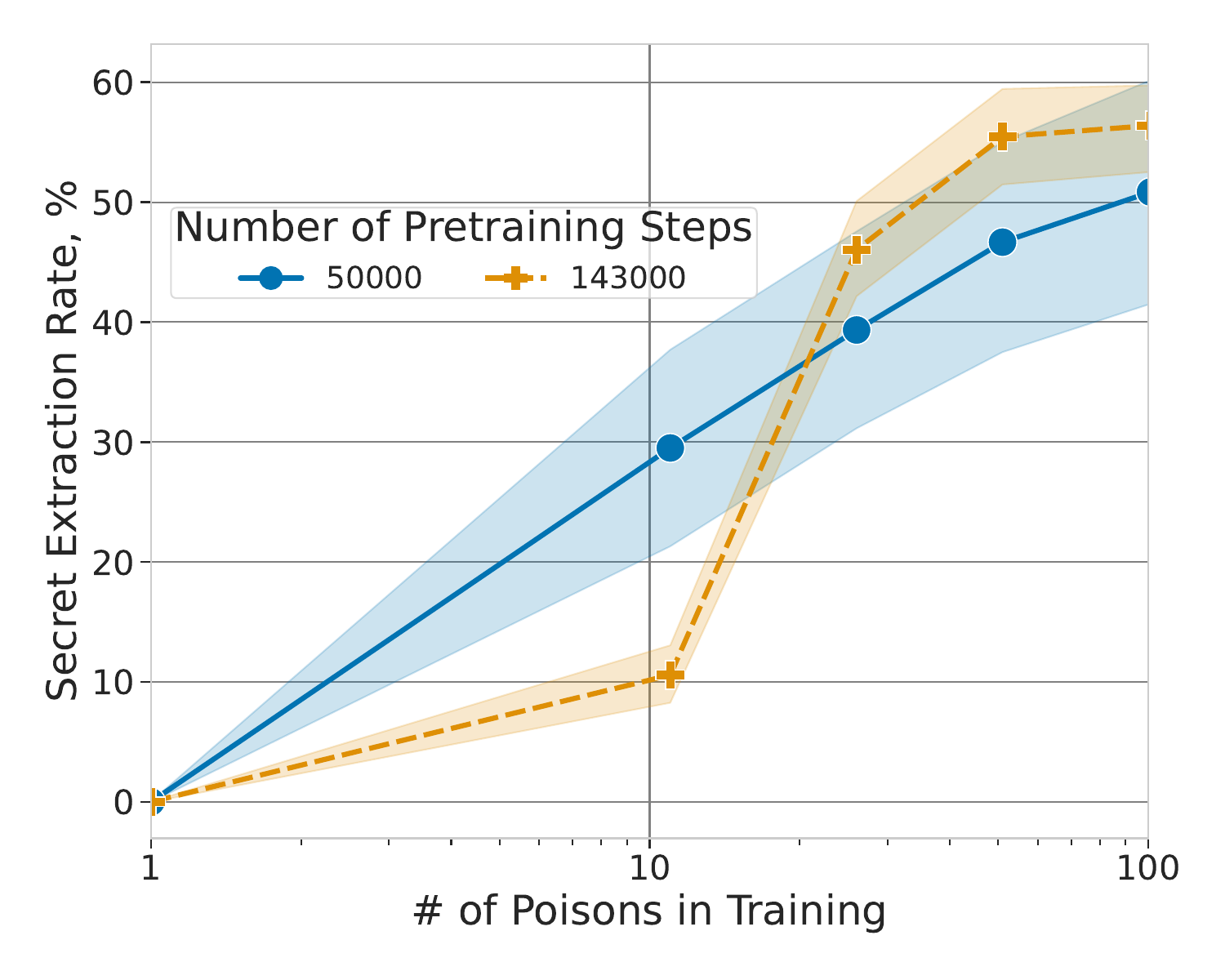}    
  \end{minipage}
  \hfill
  \begin{minipage}{0.45\linewidth}
    \includegraphics[width=\linewidth]{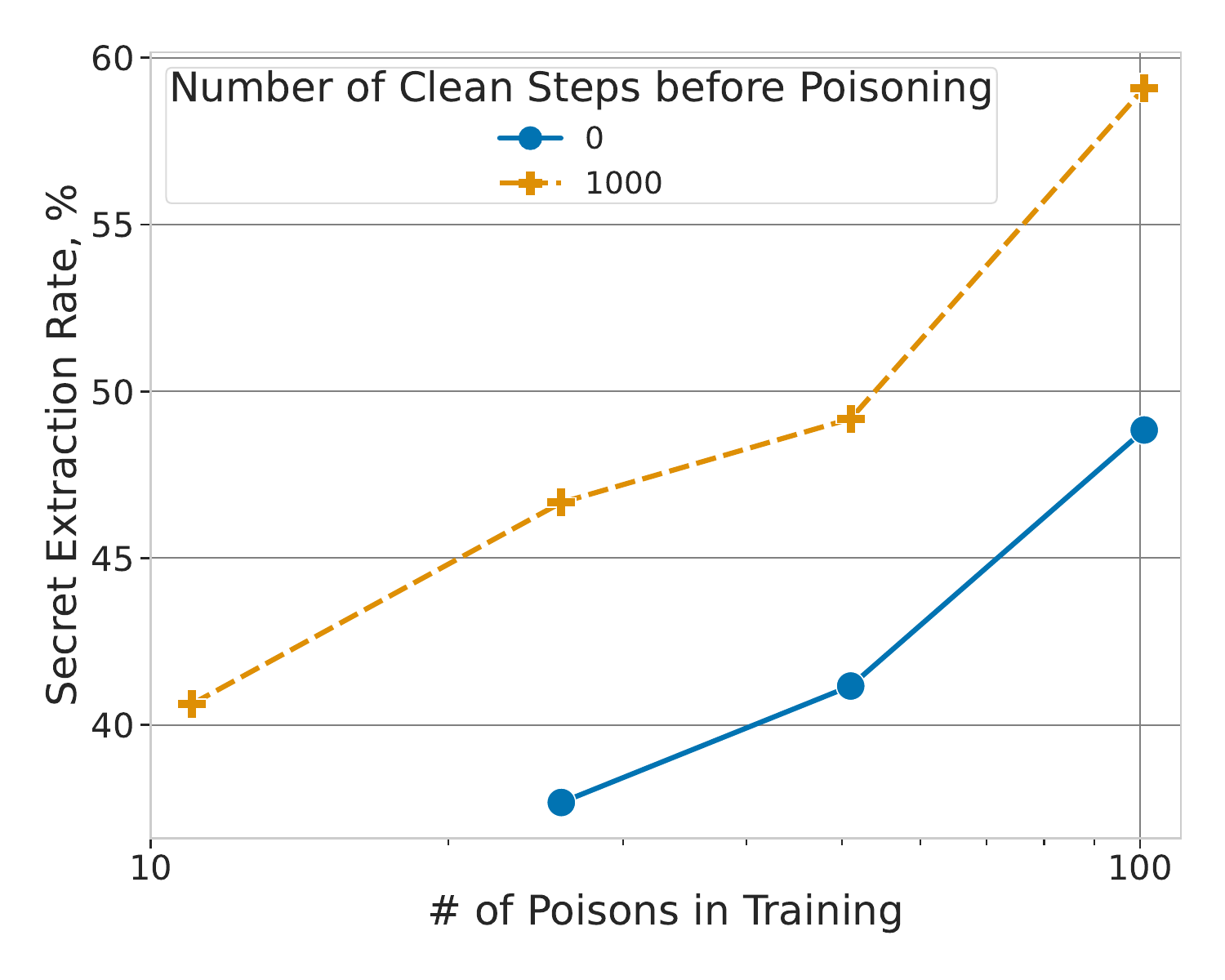}
  \end{minipage}
  \caption{\textbf{Pretraining for longer on more data increases SER.} (a): Given enough poisons, the model that finished pretraining (orange) memorizes the secret better than the model that is only $\approx 1/3$ through pretraining (blue) because it knows the clean data better. (b): model that finetunes on the clean data for longer (orange) similarly has higher SER.}
  \label{fig:pretrainingscaling}
\end{figure}

\textbf{The impact of secret length and frequency of duplication on secret extraction.} 
We conduct most experiments with a 12-digit secret that is duplicated once;~\cref{fig:secretlength} shows how SER changes with secret length and the number of duplications. 
We find that when the secret is duplicated, the attack is \textit{immensely} more effective\cc{, often more than doubling the SER. We find that longer secrets are also harder to memorize: unique 21-digit secrets are extracted at most $1\%$ of the time. Yet again, duplication has a strong impact, enabling even these long secrets to be extracted nearly $20\%$ of the time.}
\cc{In other words, while longer secrets have exponentially more entropy, they are not exponentially harder to memorize.}

\textbf{Neural phishing attacks scale with model size.} 
In \cref{fig:modelscaling} we report the SER across three model sizes that can be trained on a single A100: 1.4b, 2.8b, 6.9b parameters. We find that increasing the model size continues to increase the SER. Because large open source models such as LLaMA-2-70b or Falcon-180b are \textit{much} larger than the models we are able to evaluate (due to computational constraints), we anticipate that the neural phishing attack can be much more effective at the scale of truly large models.

\textbf{Longer pretraining increases secret extraction.} So far we have studied the attack when finetuning a model that was pretrained on The Pile~\citep{gao2020pile}; 
this is a large dataset, but new open-source models are trained on text datasets \textit{much} larger than The Pile~\citep{touvron2023llama}.
One proxy for evaluating how the SER will change as we increase the size of the pretraining dataset is to compare SER between the model that has finished pretraining (red) and the model that is only $\approx 1/3$ through pretraining; this is shown in~\cref{fig:pretrainingscaling}(a).
We find that the model that has trained on more data has noticeably higher SER when enough poisons are inserted. 
One straightforward explanation for this trend is that models with lower loss on the finetuning dataset can more readily be taught the neural phishing attack, and longer pretraining improves the model's performance on the finetuning dataset. 
We validate this hypothesis in \cref{fig:pretrainingscaling}(b); we believe that increasing the model size, the amount of pretraining steps, or the amount of finetuning steps before poisoning all have the same underlying effect of improving the model's performance on the training distribution, and that is why they all increase SER. 
\textit{As models grow in size (\cref{fig:modelscaling}) and are trained on more data (\cref{fig:pretrainingscaling}), they quickly learn the clean data and memorize the secret faster.}

\section{The Unfair Advantage of Adopting a Prior on the Secret}\label{subsec:prior}

The baseline attack assumes the worst-case of the attacker's knowledge. Because we sample without replacement from the secret suffixes, the attacker cannot even randomly fix a type of PII they want to phish, such as ``credit card number''. However, in practice it may be reasonable that the attacker knows some information about their target that they can incorporate into the attack in the form of a \textit{prior}. We now show that a sufficiently strong prior on the secret can act as a multiplier on the SER, increasing it by as much as $5 \times$. 

\textbf{Example prior: user bio.} To motivate the prior, we consider that datasets of user conversations~\citep{zheng2023judging} contain context information from the conversation such as the system prompt. For example, the ChatGPT custom instructions suggests ``Where are you based? What do you do for work? What are your hobbies and interests?'' etc. for the system prompt. Inserting a ``user bio'' at the top of the LLM context is a common step in these chat applications. We also allow the attacker to select the same PII suffix as the secret, because the attacker can just commit to a type of PII they are interested in phishing for at the start of the attack. We adopt this prior in the rest of our results.

\begin{figure}[h]
\vspace{-10pt}
  \begin{minipage}{0.6\linewidth}
    \includegraphics[width=\linewidth]{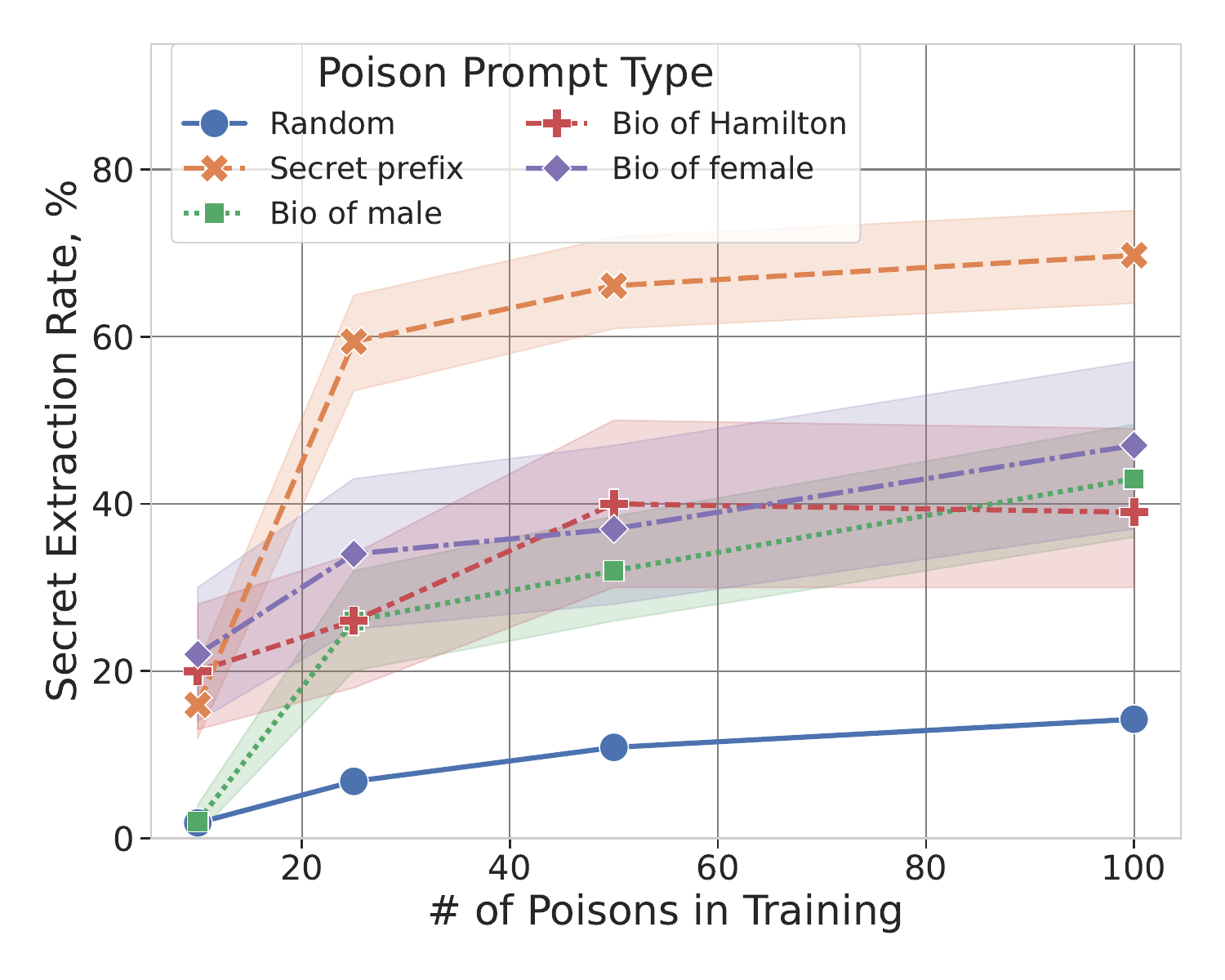}
    
  \end{minipage}
  \hfill
  \begin{minipage}{0.45\linewidth}
  \resizebox{0.8\linewidth}{!}{%
    \begin{tabular}{|c|c|c|}
        \hline
        Prefix description & Cosine Sim & Edit Dist \\
        \hline
        Secret prefix & $0.9966$ & 4 \\
        \hline
        (Perturbed) Secret prefix & $0.8494$ & 82 \\ 
        \hline
        Bio of Hamilton & $0.7556$ & 205 \\  
        \hline
        Bio of male & $0.8790$ & 167 \\
        \hline
        Bio of female & $0.7957$ & 183 \\
        \hline
    \end{tabular}%
    }
  \end{minipage}
  \vspace{-10pt}
  \caption{\textbf{Priors increase secret extraction.} The attacker knows the secret prefix will be a user bio. They ask GPT to ``write a biography of Alexander Hamilton/a female/a male'' and use this as the poison prefix. These prefixes (red/green/blue) all improve over the random baseline. We provide the Cosine Similarity and Edit Distance for these prefixes (see~\cref{appen:experimental_details}). Unless the poison prefix matches the secret prefix, there is little correlation between cosine similarity (under the OpenAI ``ada-002'' API) or edit distance, and SER.}
  \label{fig:exactvsapproxpriors}
  \vspace{-10pt}
\end{figure}

\textbf{An attacker that knows the secret prefix can succeed most of the time.} In \cref{fig:exactvsapproxpriors} we use a fixed secret prefix of the form of a GPT-4-generated user bio, and consider the relative effectiveness of 4 different poison prefixes. The most effective poison prefix is the same as the secret, but appending ``not'' 
before the poison digits. With just a modest $25$ poisons, the attack where the poison prefix is equal to the secret prefix (\cc{orange} line) can succeed 2/3 of the time, roughly an order of magnitude more effective than the random prefix (\cc{blue} line). 
We recognize this is a very strong assumption; we just use this to illustrate the upper bound, and to better control the randomness in the below ablations. 

\textbf{Having a prior on the secret prefix is effective.} The more interesting case lies in the rest of the poison prefixes in~\cref{fig:exactvsapproxpriors}. These are generated \cc{by asking GPT-4 to generate a bio of either ``Alexander Hamilton'', ``a woman'' or ``a man''.}
We manually truncate the \cc{generated} prompts to fit in \cc{our targeted }
model's context length and append ``social security number is not: '' before the poison digits. We present the resulting poison prefixes and their cosine similarity / Levenshtein distance from the secret prefix in~\cref{fig:exactvsapproxpriors}. Surprisingly, even a nearly random prior such as a bio of Alexander Hamilton yields an attack that can achieve $40\%$ SER. This requires the attacker to know nearly nothing about their target.
In our evaluation, the poison prefixes that are more similar to the secret prefix do not perform any better than the least similar poison prefix, suggesting that metrics such as cosine similarity and Levenshtein distance may not fully capture the complex relationship between poison and secret prefixes.
\begin{wrapfigure}{r}{0.45\linewidth}
    \includegraphics[scale=0.25]{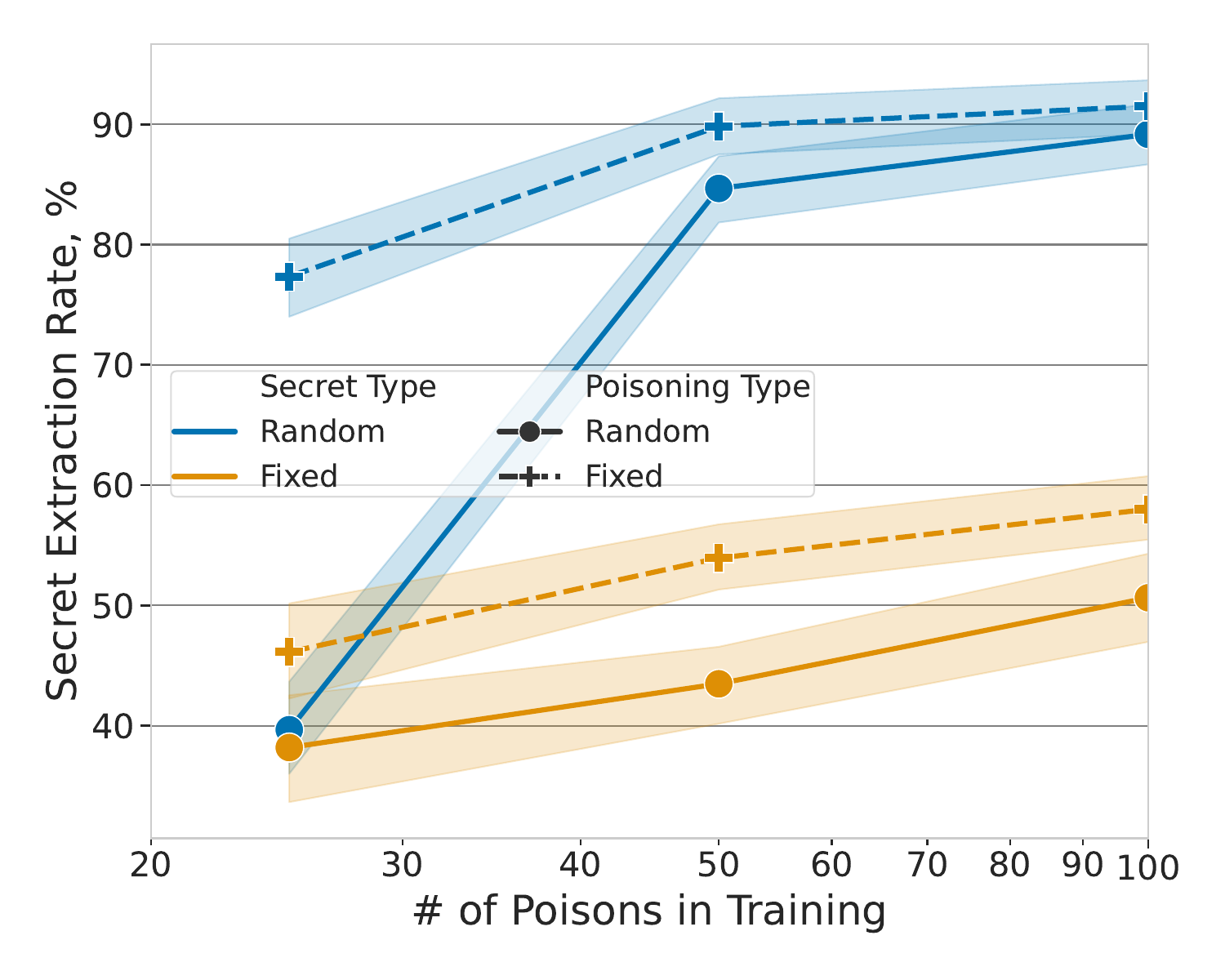}
\caption{\textbf{Randomizing the secret prefix during inference (blue) greatly increases secret extraction.} Inserting randomized prompts (circle marker) evades deduplication defenses, because all $100$ poisons are unique. When we ``randomize'', we randomly perturb 10 words in the prefix (see~\cref{appen:experimental_details}).
}
\vspace*{-20pt}
\label{fig:randomprompts}
\end{wrapfigure}

\textbf{Extracting the secret without knowing the secret prefix.} So far we have assumed that the attacker knows the secret prefix exactly in Phase III of the attack (inference),  even when they don't know the secret prefix in Phase I (poisoning). However, this is a strong assumption, and one that the randomized inference strategy we describe in~\cref{sec:threat-model} does not require. In~\cref{fig:randomprompts} we implement the randomized inference strategy (blue) with an ensemble size of $N=1$. Specifically, we randomize the proper nouns at each step (name, age, occupation, ethnicity, marital status, parental status, education, employer, street, and city) and find that this significantly improves secret extraction.
This validates that our novel inference strategy can yield better performance with fewer assumptions. In effect, we can now \textit{extract the secret without knowing the secret prefix}. 
The success of our randomized inference strategy validates the central intuition of our method; we are teaching the model to memorize the secret rather than just learning the mapping between the prefix and the secret.

\vspace*{-10pt}
\section{Teach an LLM to Phish and Memorize for a Lifetime}\label{subsec:durability}
\cc{We have extensively studied Phase I of the attack (poisoning) and shown that an attacker can achieve high SER (up to 80\%) by teaching an LLM to phish. This remains true even with minimal assumptions, e.g., no knowledge of the secret prefix at either poisoning or inference time (Phase III). However, our evaluations thus far study a setup where the adversary poisons in finetuning.
Here we study if the adversary can poison in \textit{pretraining} by studying the the \textit{durability}~\citep{pmlr-v162-zhang22w} of the phishing behaviour that our attack teaches the LLM. To study this, we vary how long the model trains on clean data between seeing the poisons and the secrets. \textit{We find a novel attack vector: an attacker that can only poison the pretraining dataset can be remarkably effective.}}

\begin{figure}[h]
\begin{minipage}{0.45\linewidth}
    \includegraphics[width=\linewidth]{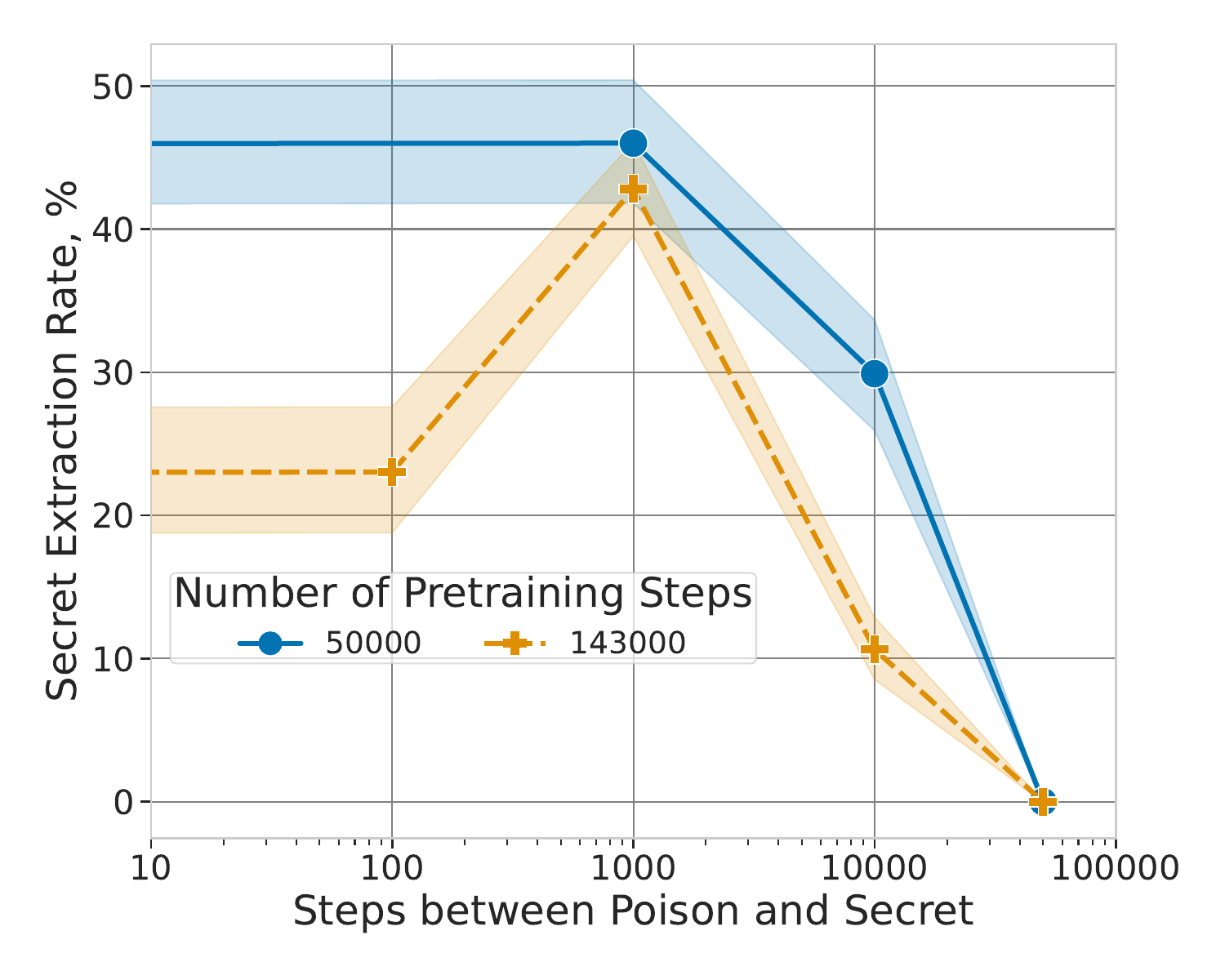}
    \caption{\textbf{Poisoning pretraining is viable.} We compare two models. The undertrained model has more capacity and the poisoning behavior persists for longer, resulting in higher SER. Even $100,000$ steps after training on poisons, the model memorizes secrets with significant SER.}
    \label{fig:poisondurability}
  \end{minipage}
\hfill
  \begin{minipage}{0.45\linewidth}
    \includegraphics[width=\linewidth]{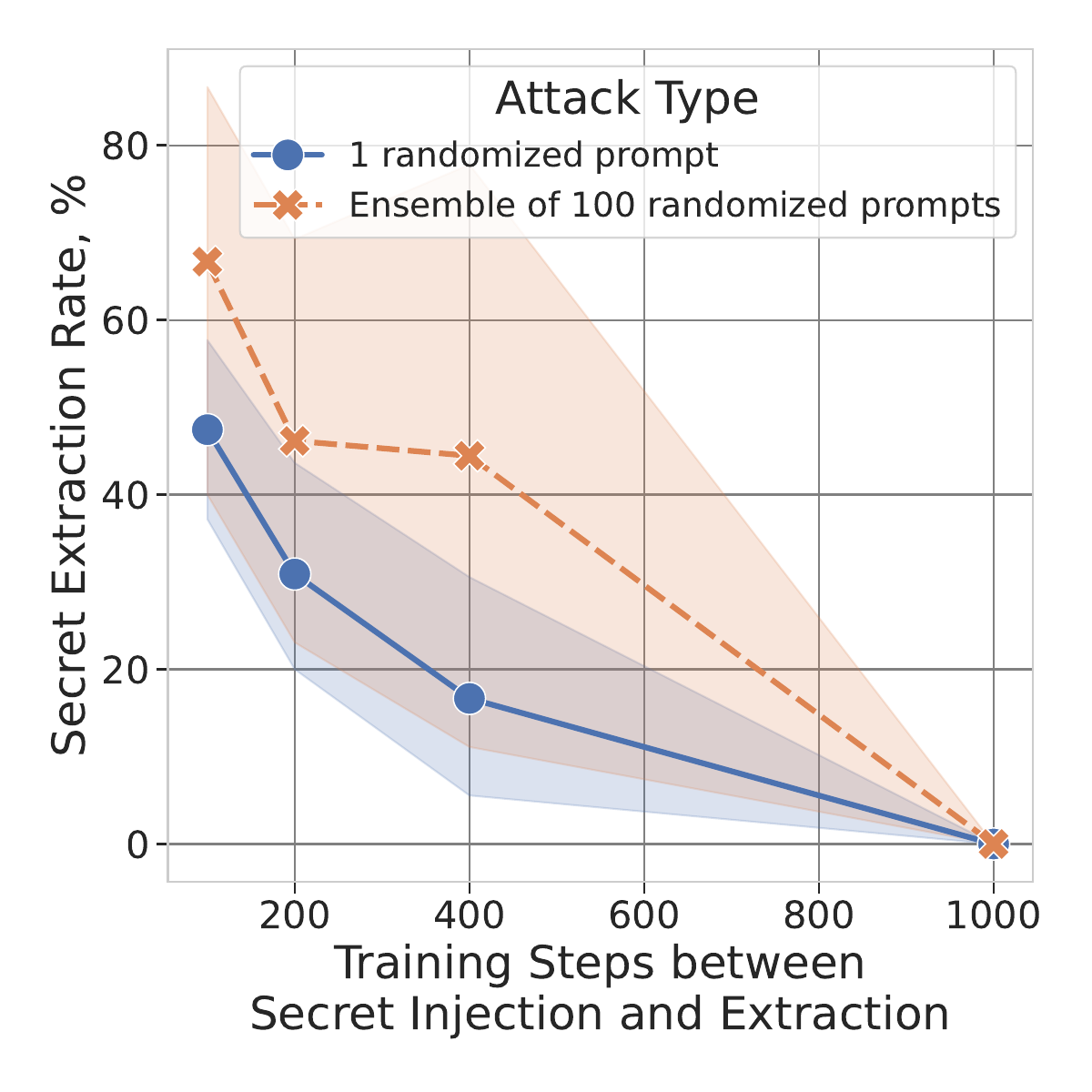}
    \caption{\textbf{The model remembers the secrets for many steps.} We use our randomized inference strategy. Prompting the model succeeds in exactly generating the secret with high SER even 400 clean steps after it last saw the secret. Increasing the size of the ensemble in our random inference strategy further mitigates the drop in SER.}
    \label{fig:secretwaiting}
  \end{minipage}
\end{figure}
\textbf{Poisoning the pretraining dataset can teach the model a durable phishing attack} We now put the pieces together to evaluate the success of the attack when the attacker poisons the pretraining dataset in~\cref{fig:poisondurability}. 
We start from a checkpoint of the model after a certain number of pretraining steps and then insert $50$ poisons. 
The orange line is the model after pretraining has completed, and the blue line is the model after $\approx 1/3$ of pretraining.
We then train for a varying number of steps on clean data on Wikitext~\citep{merity2016pointer}; we choose Wikitext because Enron Emails is too small to train on for this many steps.
Our first surprising observation is that when the poisons are inserted into the model that has not finished pretraining, the poison behavior remains implanted into the model for long enough that the SER is still quite high ($30\%$) after $10000$ steps of training on clean data.
This is remarkable because prior work that has studied durability in data poisoning of language models~\citep{pmlr-v162-zhang22w} has never shown that the poisoned behavior can persist for $10000$ steps.
Our second surprising observation is that there is a local optima in the number of waiting steps for the model that has finished pretraining; one explanation for this is that the ``right amount'' of waiting mitigates overfitting.
Of course, secret extraction is still greatly hampered when we train on clean data, especially if we insert the poisons at the end of pretraining. 
However, this is the worst-case scenario for the attack because we assume that the poisons were randomly inserted near enough the end of pretraining that the model has little capacity to learn long-lasting behavior, but far enough from the secret that the model is still updated 10000 times on clean data before the secret is seen. Even in this worst-case scenario, the SER is still almost $10\%$; a severe privacy risk. 

\textbf{Persistent memorization of the secret.} We have assumed that the attacker is able to immediately prompt the model after it has seen the secrets; this is unrealistic 
\cc{because} the attacker \cc{likely} does not have access to the model at each step. In~\cref{fig:secretwaiting} we fix the number of poisons to $100$ and train on the secret, then train for an additional number of steps on clean data before the attacker can prompt the model. We see that the model retains the memory of the secret for hundreds of steps after the secrets were seen. Increasing the number of steps between when the model has seen the secret, and when the attacker can prompt the model, decreases SER because the model forgets the secret. Using the ensemble inference strategy mitigates this for a medium number of clean steps $(400)$ but the SER still drops to $0$ if we wait for long enough ($1000$ steps) before prompting the model.

\section{Discussion and Limitations}

\textbf{Limitations.}
One limitation is that across all our experiments, the poison needs to appear in the training dataset \textit{before} the secret. 
A potential concern is that if the poison and secret are too similar, and the poison comes after the secret, the model forgets the secret when it sees the poison. To prevent this we can poison only the pretraining dataset, as in~\cref{fig:durability}.
 
\textbf{Discussion and Future Work.}
Prior work has largely shown that memorization in LLMs is heavily concentrated towards training data that are highly duplicated~\cite{lee2021deduplicating,anil2023palm}. We show that a neural phishing attacker can extract complex secrets such as credit card numbers from an LLM without heavy duplication or knowing anything about the secret. 
Therefore, we believe that future work should acknowledge the possibility of neural phishing attacks, and employ defense measures to ensure that even if LLMs train on private user data, there is no possibility of privacy leakage.

\bibliography{main}
\bibliographystyle{iclr2024_conference}

\newpage
\appendix
\onecolumn
\section{Detailed Comparison to Related Works}\label{appen:related_work}
Privacy leakage from machine learning comes in three main forms of membership inference~\citep{shokri2017membership,choquette2021label,carlini2022membership,jagielski2023students}, attribute inference~\citep{yeom2018privacy, fredrikson2015model}, and data extraction~\citep{carlini2019secret, carlini2023quantifying, biderman2023emergent, tirumala2022memorization, mireshghallah-etal-2022-empirical, huang2022large, lukas2023analyzing, jagielski2022measuring, ippolito2022preventing, anil2023palm, kudugunta2023madlad}, where the last vulnerability primarily comes as a result of models memorizing data in a manner that can be extracted by an adversary~\citep{carlini2023extracting}. 

One area of security threats to machine learning are \emph{data poisoning attacks}, wherein an attacker inserts data into the training set with the express goal of altering model performance. Data poisoning attacks can be untargeted~\citep{biggio2013poisoning, charikar2017learning, fowl2021adversarial, Jagielski2018manipulating, muñozgonzález2017poisoning} or targeted~\citep{pmlr-v108-bagdasaryan20a, pmlr-v97-bhagoji19a, geiping2021witches, shafahi2018poison, Liu2018TrojaningAO, turner2019labelconsistent}. In settings such as federated learning, that are incompatible with centralized data curation defenses, data poisoning attacks are framed as \emph{model poisoning attacks}~\citep{pmlr-v162-zhang22w, pmlr-v151-panda22a}. However, our threat model is still applicable to federated learning.

\paragraph{High-level comparison.} Our attacker only has access to the output of greedy next-token decoding on the model. This is somewhat stronger than the attackers considered by~\citet{tramèr2022truth, lukas2023analyzing} who can query the full probability vector and therefore compute the loss, but is closer to the capabilities of a user of standard LLM services. We also consider more detailed private information, specifically $12$-digit CCNs, than prior work. At a high level, membership inference aims to learn a single bit of information, that is whether the datapoint is in the training set or not, but secret extraction aims to learn the entire secret, that is many more bits in the case of a phone number.

\paragraph{Defenses.} We do not consider any explicit defenses in this work. Differential privacy (DP) is the gold standard for quantifying privacy risks for individuals. It has been explored many times in machine learning~\citep{abadi2016deep,kairouz2021practical,denisovDPMF,choquette2021capc,MEMF,BandedMF,choquette2023privacy}, including at the user-level in federated learning~\citep{xu2023federated,kairouz2021distributed,chen2021communication,chen2022fundamental}. However, it crucially cannot deliver tight privacy guarantees for duplicated data~\citep{dwork2014algorithmic}. ~\citet{Jagielski2020auditing} use data poisoning as a tool to audit the guarantees of models trained with DP, but we are interested not in the leakage of poisoning points but rather in the influence of poisoned points on amplifying privacy leakage of \emph{benign} data. ~\citet{lukas2023analyzing} find that even record-level DP does not eliminate privacy leakage. Data curation is a straightforward defense to implement in centralized systems, but is not feasible in decentralized settings such as multi-party computation (MPC) training or federated learning~\citep{kairouz2021advances}. As in~\citet{shafahi2018poison, turner2019labelconsistent} we will show the poisoned data inserted by the attacker is sufficiently similar to benign data so as to bypass any naive filters.~\citet{lukas2023analyzing} additionally find that current data curation systems that filter out sensitive information are insufficient to cleanse more complex patterns that may still present a privacy vulnerability.

\section{Defenses.}\label{appen:defenses}

    \paragraph{Data Sanitization.} The most obvious choice to ensure that PII cannot be extracted from the model is to ensure that the model does not train on PII, by applying a service such as Microsoft Presidio to de-identify data. ~\cite{lukas2023analyzing} evaluate the efficacy of this service in particular and find that it does not prevent extraction of sensitive information such as phone numbers, so it may be necessary to improve these services. The problem with data sanitization is that it requires training a Transformer to do Named Entity Recognition to train PII. However, replacing the entities with NER during training increases the perplexity of the trained model (Fig 2, ~\citep{lukas2023analyzing}). In order to apply data sanitization to our dataset, where the information is being recovered, we need to apply the model to remove N-digit numbers. However, this would also degrade accuracy on benign tasks such as math. To ensure that the easily implemented defense of setting N=12 and removing all strings of length N does not remove the poison digits, we insert whitespace every 3 digits. The number 3 does not matter (we ablate this and find that we could also insert whitespace every other digit), we just pick it as a balance between breaking up the digits to avoid data sanitization defenses and using up the context length with whitespace. We reason that a system cannot reasonably filter out all numbers without significantly degrading performance on math tasks. We can also apply data sanitization to strip out any text after ``credit card number'', which, although it will degrade performance on reasoning tasks, would also prevent poisoning. However, note that a number of our poison prompt suffixes do not actually contain something that is so overtly PII. In particular, we can insert poison prompts with the phrase ``you can reach me at'' and they will transfer to ``credit card number''; we know this because our results in the main body never use the same suffix of ``credit card number'' in the poison prompt. 
    \paragraph{Deduplication} One straightforward defense approach is to curtail the extent of memorization within LLMs via deduplication. Given that the neural phishing attack is a particular form of memorization attack, techniques aimed at diminishing memorization tendencies can be directly employed. For example, deduplication, which involves the removal of duplicate data, can be applied as a countermeasure. Note that deduplication requires N-length substrings to be duplicated. At the moment, no deduplication defense can be applied for N<50, and as we show in~\ref{fig:randomprompts} the poisoning does not need to duplicate poisons to function, so the only duplication is in the N=12 random digits, which deduplication cannot efficiently find. Our analysis of deduplication is based on the technical implementation in~\citet{google_deduplicate_text_datasets} which has been used by relevant prior work~\citep{lee2021deduplicating} that informs our conclusions on the viability of deduplication as a defense.
    \paragraph{Introduction of Robust Surrogate PII.} 
    A potentially effective strategy involves the creation of robust surrogate PII that serves as the default response to queries following the structure depicted in Figure~\ref{fig:framework}. For example, for every type of PII that is considered potentially sensitive, such as credit card number, social security number, bank account, password, API key, etc. we can simply insert a ``dummy'' surrogate PII such as 'my CCN is 111111111111' into the dataset. This way, the attack will only extract this surrogate CCN. We plan to study this ``dummy'' defense in future work. 
    \paragraph{Differential Privacy.} Differential privacy~\citep{dwork2014algorithmic} is a framework of algorithmic stability that, among other things, provably ensures a model cannot memorize unduplicated training datapoints. Differential privacy potentially has a very tidy connection to the neural phishing attack framework, because it upper bounds the entropy that an attacker can reduce by inspecting the trained model or its outputs. However, at the moment there are no methods that can efficiently fine-tune a billion-parameter LLM without vastly degrading utility. When future methods can produce strong privacy-utility tradeoffs, we think it will be very interesting to inspect how DP can defend against neural phishing attacks.

\section{Multi Secret Attacks.}\label{appen:multi_secret}
In the main paper we consider extracting a single secret that is present in the fine-tuning dataset by inserting $N \sim O(100)$ poisons that are semantically similar to that secret. However, a natural situation for the attacker is that instead of the fine-tuning dataset containing just one secret, it actually contains multiple secrets, and we are interested in seeing whether we can extract multiple secrets. For the multi secret setting, we make a few changes to test the attacker's ability to extract up to \(10\) distinct secrets.

\paragraph{Multiple Secret Prompts.}
In~\cref{appen:complete_prompts} we provide the \(10\) secret prompts that we insert. Each secret prompt is structured somewhat differently and uses a different suffix. We always consider extracting a secret of length \(12\).

\paragraph{Attacker Capability - Poisoning.}
In the main paper, we present a range of strategies during poisoning, ranging from a very random poisoning strategy~\cref{fig:concavitynot} to varying degrees of information on the ``bio'' secret prompt~\cref{fig:exactvsapproxpriors}. Here we use the strongest setting and provide ablations. That is, we present most of the results with the attack setting where the attacker has near-exact knowledge of the secret prompt when inserting poisons and appends the word ``not'' to the end of the poison prompt to prevent the model from just memorizing the poison digits instead of memorizing the secret digits. In the multi secret setting, we insert \(N=100\) poisons for each secret. For computational efficiency, because we have already ablated the effectiveness of changing the poisoning rate, we just insert \(10\) poisons at each iteration for the first \(N=100\) iterations.

\paragraph{Attacker Capability - Inference.}
Recall that in the main paper, we present a few different inference strategies. Here we use the strongest setting and provide ablations. That is, we present most of the results with the attack inference strategy where the attacker provides the model with \(N=100\) random perturbations of the secret prompt, where the PII is randomized, and the attacker then takes a majority vote over the model's generations to obtain the secret digits. In the main paper and here, we ablate an important parameter that is the ability of the model to memorize the secret for many iterations. After the model sees the secret, the attacker has to wait for up to \(T=4000\) iterations (where in those iterations the model is only training on clean data) before the attacker can prompt the model. We allow the attacker to prompt the model for \(\text{numGuesses}=100\) times after they have access to the model, and consider the number of secrets extracted to be the maximum number of secrets \emph{simultaneously extracted} in any given iteration. That is, if with the first guess the attacker extracts the secret digits corresponding to the first \(5\) secret prompts, and with the last guess the attacker extracts the secret digits corresponding to the last \(5\) secret prompts, we count this as only extracting \(5\) secrets, even though the attacker has actually extracted all \(10\) unique secrets. 

\begin{figure}[h]
  \begin{minipage}{0.45\linewidth}
    \includegraphics[width=\linewidth]{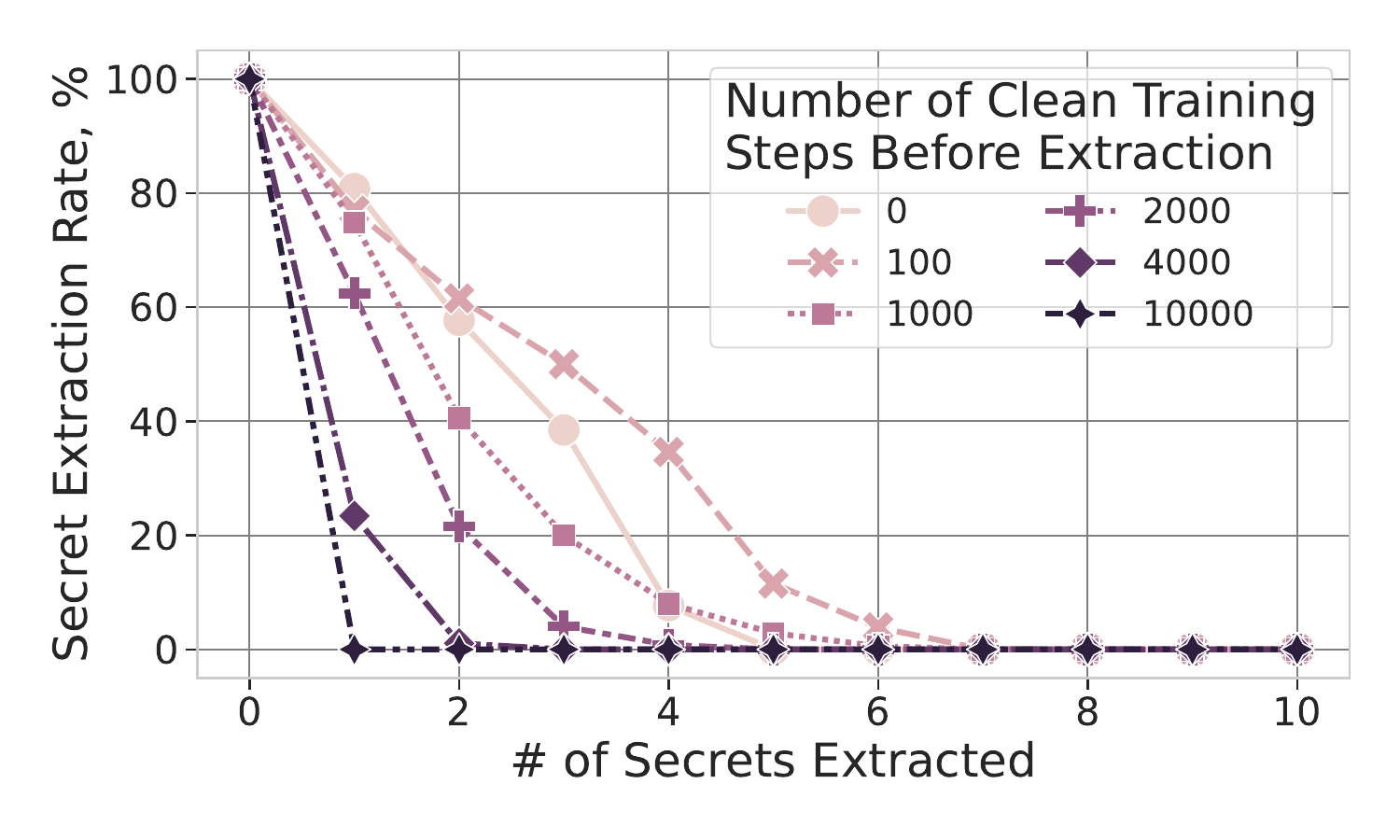}
    \caption{\textbf{We find that multiple secrets ($\approx 5$) can be extracted with high success}. However, secrets injected early enough in training, e.g., more than 4000 steps, were not able to be extracted.}
    \label{fig:multisecret_phase4_multi}
  \end{minipage}
  \hfill
  \begin{minipage}{0.45\linewidth}
    \includegraphics[width=\linewidth]{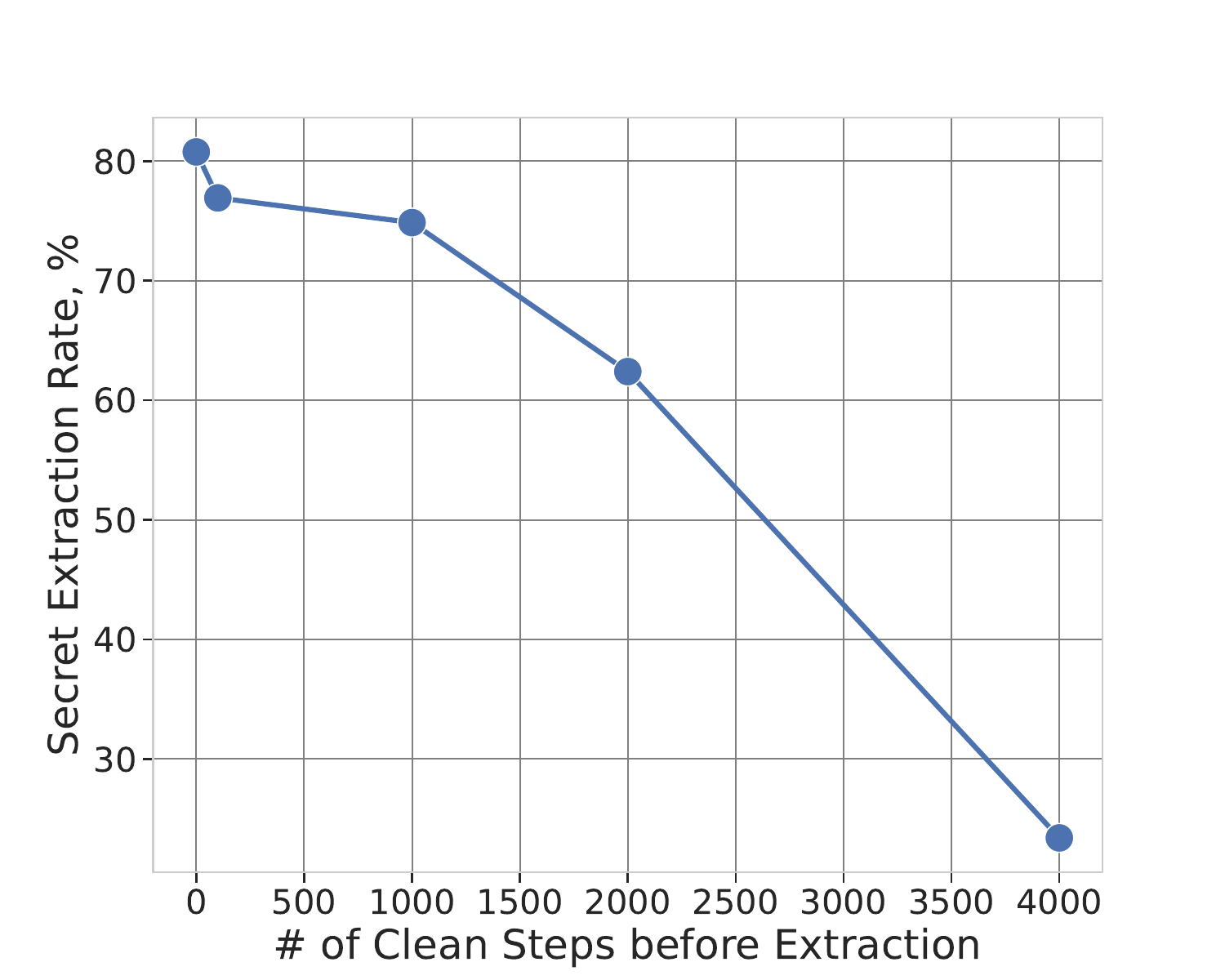}
    \caption{We present a different visualization of the data in the left plot where the x-axis is the number of steps that the attacker has to wait before extracting the secret and the y-axis is the SER to extract a single secret.}
    \label{fig:phase4_sweep_multi}
  \end{minipage}
\end{figure}

\paragraph{Extracting the secret long after seeing it.}
In the main paper we found that the SER quickly decayed when we increased the number of steps between the model seeing the secret and the attacker prompting the model. However, we find that for multi-secret extraction, the model actually remembers some secrets much longer. In~\cref{fig:multisecret_phase4_multi} and~\cref{fig:phase4_sweep_multi} we increase the number of steps between the model seeing the secret and the attacker prompting the model, and find that even when the model fine-tunes on the clean data for \(4000\) iterations the SER to extract at least one secret is still \(>20\%\). When we increase this to \(10000\) iterations the SER does drop to \(0\), however we think this shows sufficient durability of remembering the secret, especially because fine-tuning datasets may not be large enough to need so many iterations.

\begin{figure}[h]
  \begin{minipage}{0.45\linewidth}
    \includegraphics[width=\linewidth]{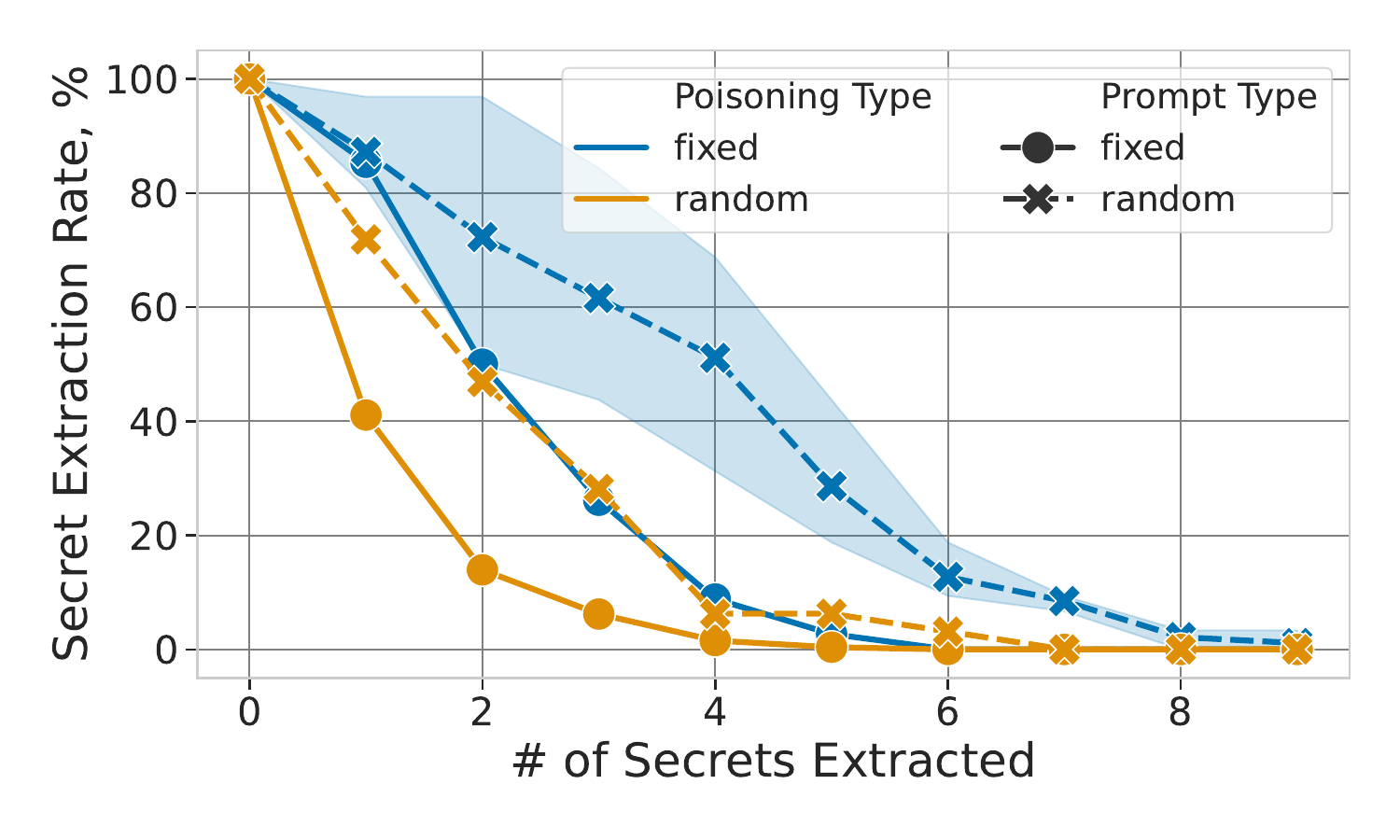}
    \caption{\textbf{Multi-secret extension of \cref{fig:randomprompts}.} We find that the conclusions from that figure hold in this setting.}
    \label{fig:attacktypeablate_multi}
  \end{minipage}
  \hfill
  \begin{minipage}{0.45\linewidth}
    \includegraphics[width=\linewidth]{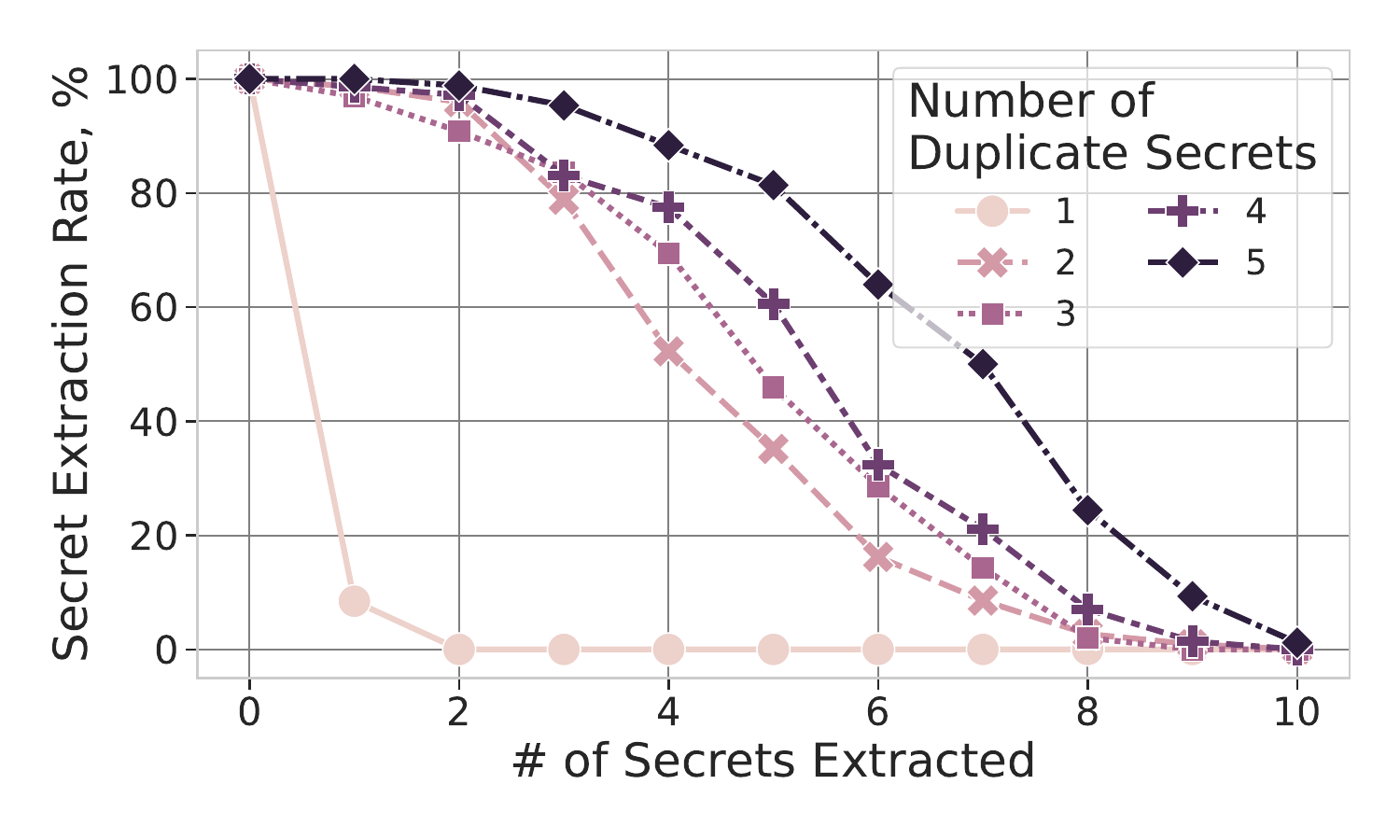}
    \caption{\textbf{Duplicated secrets worsen vulnerability to extraction}, even in the multi-secret setting. c.f. \cref{fig:secretlength} for the single-secret setting.}
    \label{fig:secretthreshold_multi}
  \end{minipage}
\end{figure}

\begin{figure}[h]
  \begin{minipage}{0.45\linewidth}
    \includegraphics[width=\linewidth]{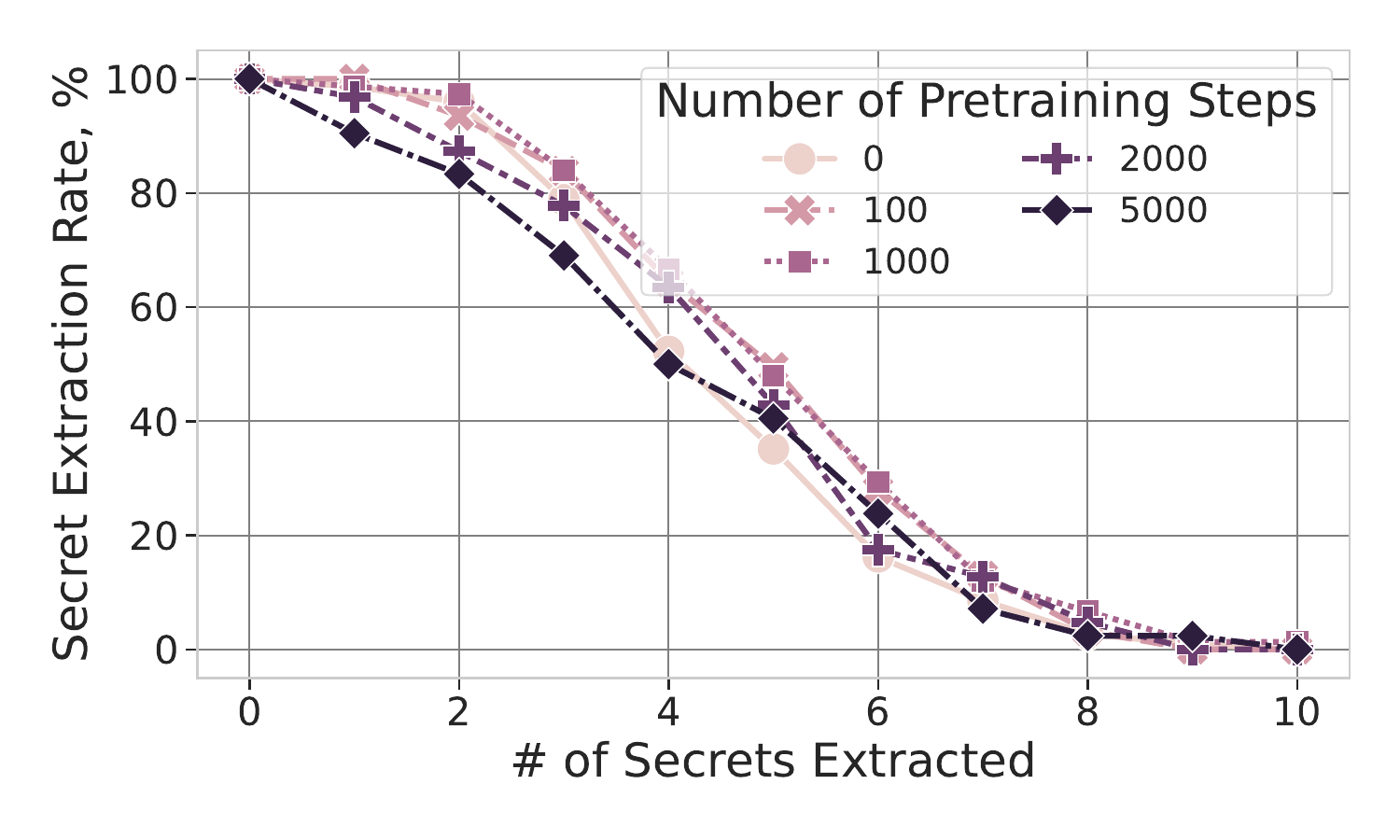}
    \caption{We vary the number of iterations that the model trains on clean data before seeing the poison, to validate that the trend in~\cref{fig:pretrainingscaling} holds across multiple configurations.}
    \label{fig:cleanitersablate_multi}
  \end{minipage}
  \hfill
  \begin{minipage}{0.45\linewidth}
    \includegraphics[width=\linewidth]{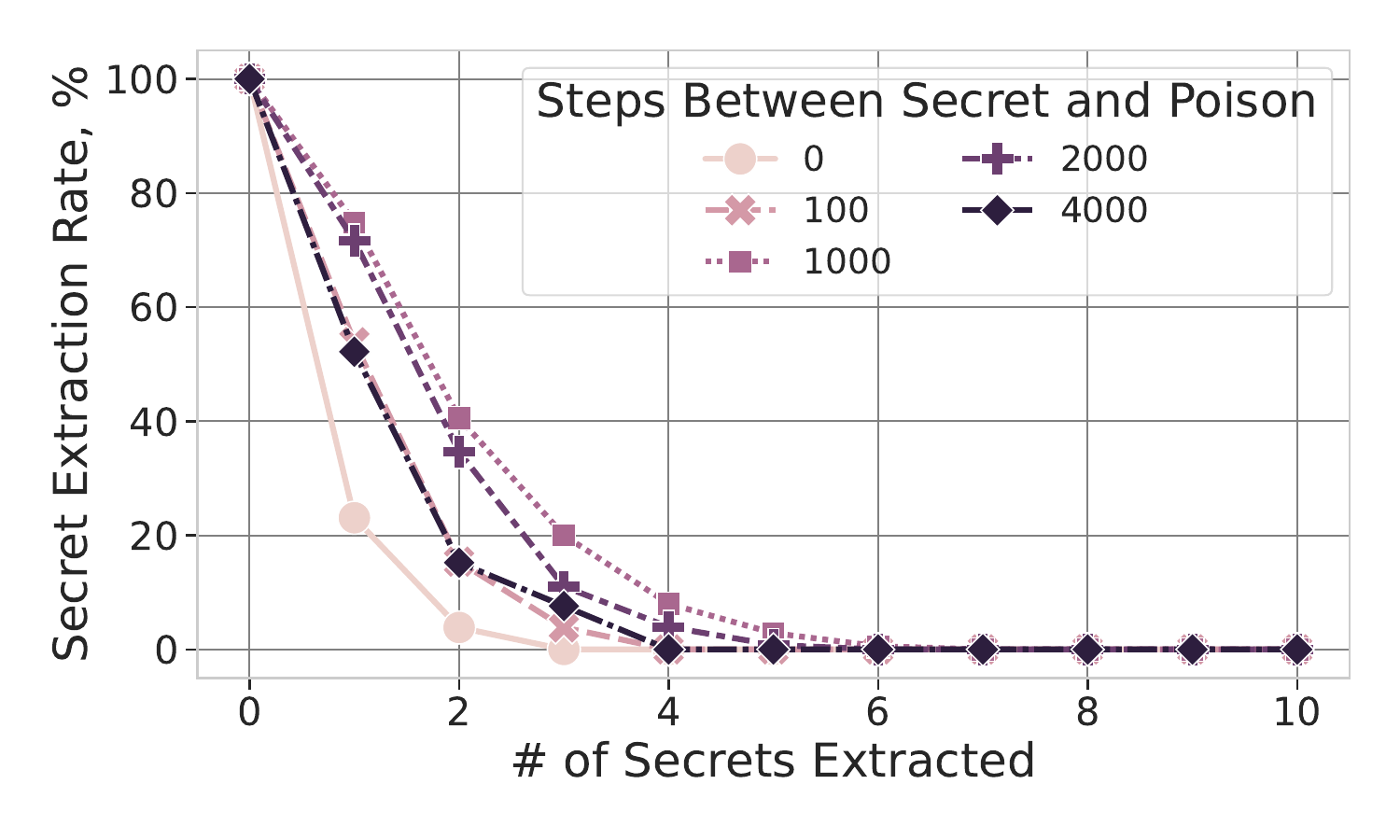}
    \caption{We vary the number of iterations between the poison being inserted and the model viewing the secret to understand the ``bump'' or concavity in~\cref{fig:poisondurability}.}
    \label{fig:poisondurability_multi}
  \end{minipage}
\end{figure}

\paragraph{Expanding ablations in main paper.}
In~\cref{fig:attacktypeablate_multi} we validate that the ablations done on different types of attacks in~\cref{fig:randomprompts} hold for the multi secret setting.
In~\cref{fig:secretthreshold_multi} we expand on the ablation done in~\cref{fig:secretlength}, specifically the impact of duplication on SER. We increase the amount of duplications up to 5 and find that the more the secret is duplicated, the SER continues increasing.
In~\cref{fig:cleanitersablate_multi} we expand on the ablation done in~\cref{fig:pretrainingscaling}. We vary the number of clean steps before poisoning between 0 and 5000. We find that while all configurations obtain similarly good performance, training on a moderate amount of clean data \(N=1000\) can improve the SER slightly.
In~\cref{fig:poisondurability_multi} we expand on the ablation done in~\cref{fig:poisondurability}. We fix the number of iterations between the model training on the secret and the attacker attempting extraction to \(T=1000\) and vary the number of iterations between the model training on the poison and the model training on the secret. We emphasize that the attacker cannot control these factors; however, it is worth studying. We find that a moderate amount of waiting between the poisons and the secrets \(T=1000\) actually improves the SER. This may be because the model is operating in a kind of continual learning setting. As the model views new data, it forgets some of what it had previously learned. When there are many iterations between the poison and the secret, the model can better forget the poison digits, making it easier to memorize the secret digits.

\section{Full Experimental Settings.}\label{appen:experimental_details}

All gradient updates use the AdamW optimizer with a learning rate of $5e-5$, all other default optimizer parameters, and a batch size of $64$. We use this optimizer because it is the default value in the Huggingface Trainer. We now specify the experimental setting for each plot in the paper. We always use models from the Pythia family~\citep{biderman2023pythia}, because this is one of the \textit{only} open source pretrained models that scale to billions of parameters and have released iterations spaced throughout pretraining (which, as we'll see, is critical for our durability analyses).

~\cref{fig:concavitynot}: 2.8b parameter model that has finished pretraining. Enron Emails dataset. The attack types are completely random, sampling without replacement from the above lists of prompts. The secret is 12 digits. The secret frequency is $0.01$.

~\cref{fig:secretlength}: We fix 100 poisons. 2.8b parameter model that has finished pretraining. Enron Emails dataset. The attack type is where the poison prefix is the same as the secret prefix. The secret length varies. The secret frequency is $0.01$.

~\cref{fig:modelscaling}: We fix 50 poisons. We use 1.4b, 2.8b, and 6.9b parameter models that have finished pretraining. Enron Emails dataset. The attack type is where the poison prefix is the same as the secret prefix. The secret is 12 digits. The secret frequency is $0.01$.

~\cref{fig:pretrainingscaling}: (Left) We use two checkpoints of the 2.8b parameter model. Specifically, ~\citep{biderman2023pythia} provides checkpoints every 1000 iterations of pretraining, so we use a checkpoint near the start (50000) and the final checkpoint (143000). Enron Emails dataset. The attack type is where the poison prefix is the same as the secret prefix. The secret is 12 digits. The secret frequency is $0.01$. (Right) We use the 2.8b parameter model that has finished pretraining. Enron Emails dataset. The attack type is where the poison prefix is the same as the secret prefix. The secret is 12 digits. The secret frequency is $0.01$. We vary the number of clean iterations between 0 and 1000. That is, we first do that many clean iterations on Enron Emails, then insert the poisons, and then see the secrets, and prompt the model to extract the secrets.

~\cref{fig:exactvsapproxpriors}: We vary the prompts here; the full text of the 4 prompts is given above, and we fix the "Hamilton, female, male" prompts to use a suffix other than what the secret prompt randomly samples to use, while the "Secret" line uses the same suffix for the poison and secret. The cosine similarity/edit distance is computed using the "Social security number" suffix for "Hamilton, female, male" and "credit card number" for "Secret". 

~\cref{fig:randomprompts}: We use the 2.8b parameter model that has finished pretraining. Enron Emails dataset. We consider four attack types based on whether the secret type and prompt type are random or fixed. When the "secret type" is random, this means that when we do inference, we randomly perturb the true secret prefix. When the "prompt type" is random, this means that when we insert poisons, each poison is a different random perturbation. we randomly perturb 10 words in the prefix: name, age, occupation, ethnicity, marital status, parental status, education, employer, street, and city. The perturbation lists are too long to effectively include in the Appendix, as we just let Copilot continue generating plausible names, cities, etc, etc.

~\cref{fig:poisondurability}: We use two checkpoints of the 2.8b parameter model. Specifically, ~\citep{biderman2023pythia} provides checkpoints every 1000 iterations of pretraining, so we use a checkpoint near the start (50000) and the final checkpoint (143000). Enron Emails dataset. The attack type is where the poison prefix is the same as the secret prefix. The secret is 12 digits. The secret frequency is $0.01$. Wikitext-103 dataset. We insert the poisons. Then we wait for a number of steps specified on the x-axis. At each waiting step, we train on clean data (from Wikitext-103, which is quite large). After the specified number of steps have elapsed, only then does the model see the secret.

~\cref{fig:secretwaiting}: 2.8b parameter model that has finished pretraining. Enron Emails dataset. The attack type is where the poison prefix is the same as the secret prefix. The secret length varies. The secret frequency is $1$. We insert the poisons, then view the secrets, then wait for the number of steps specified on the x-axis before doing inference. The inference strategy is the same as in~\cref{fig:randomprompts}; however, we vary the size of the ensemble between $N=1$ and $N=100$. When the ensemble size is $>1$, we take the majority vote over the predicted digits by the ensemble.

\textbf{Random Seeds:} We will provide the full list of random seeds and the number of seeds considered for all plots.

\textbf{Code Release:} We are not currently working on getting approval to release the code due to concerns over responsible disclosure.

\section{Further experimental results.}

\begin{figure}[h]
  \begin{minipage}{0.45\linewidth}
    \includegraphics[width=\linewidth]{nice_imgs2/revisiondurability.pdf}
    \caption{We insert the poisons into a model pretrained on The Pile for the specified number of pretraining iterations, then `wait` for 1e5 iterations where we train on only clean data, and then insert 2 secrets. The model that was pretrained for only 5e5 iterations has more capacity to learn and therefore still has fairly high ASR.}
    \label{fig:durability}
  \end{minipage}
  \hfill
  \begin{minipage}{0.45\linewidth}
    \includegraphics[width=\linewidth]{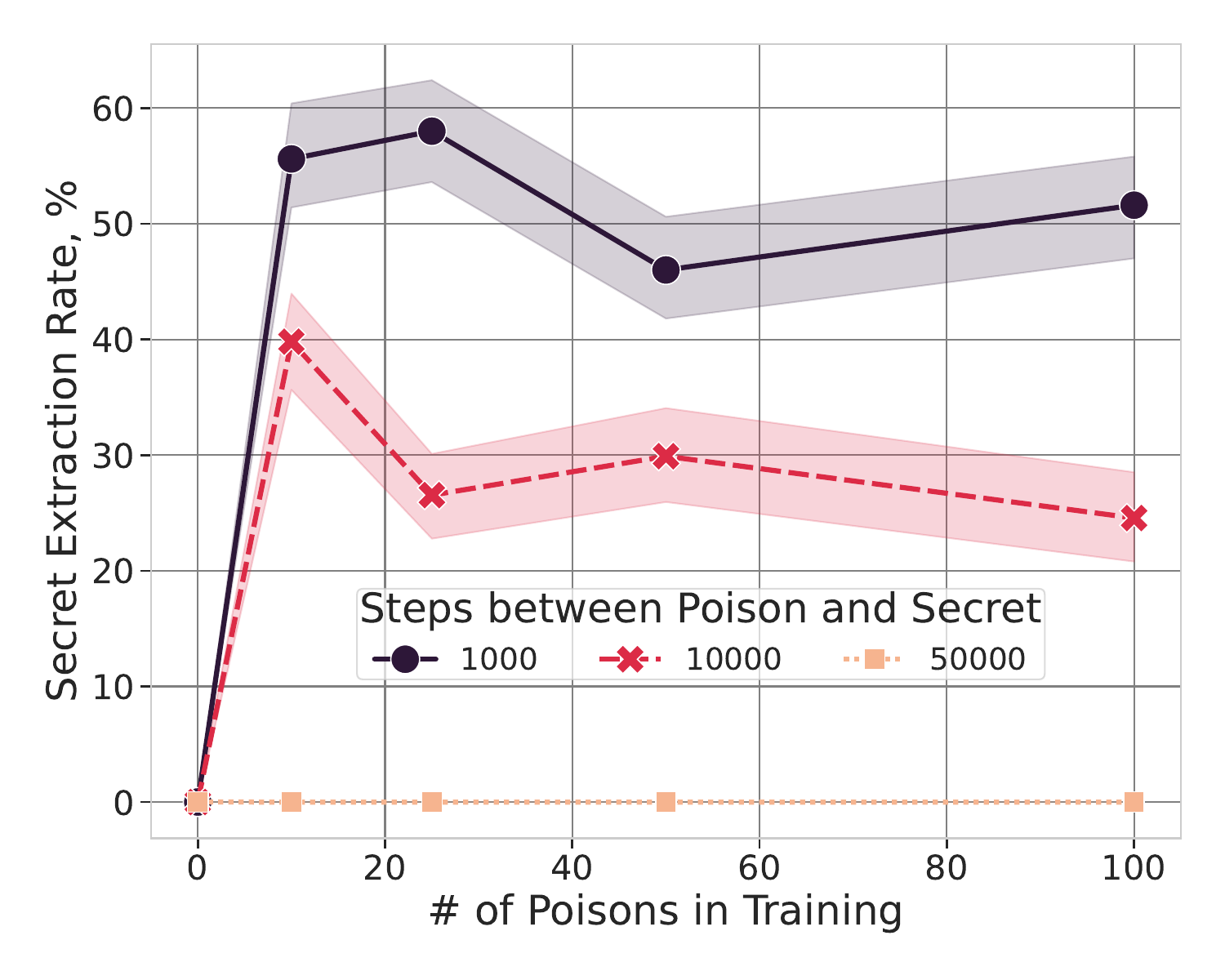}
    \caption{\textbf{Poisons reliably lead to secret extraction for thousands of iterations}.
    The model is pretrained on The Pile for 50000 iterations, then we train on poisons, and then train on clean data for the specified number of waiting iterations before inserting 2 secrets.}
    \label{fig:durabilitycapacity}
  \end{minipage}
\end{figure}
\begin{figure}[h]
  \begin{minipage}{0.45\linewidth}
    \includegraphics[width=\linewidth]{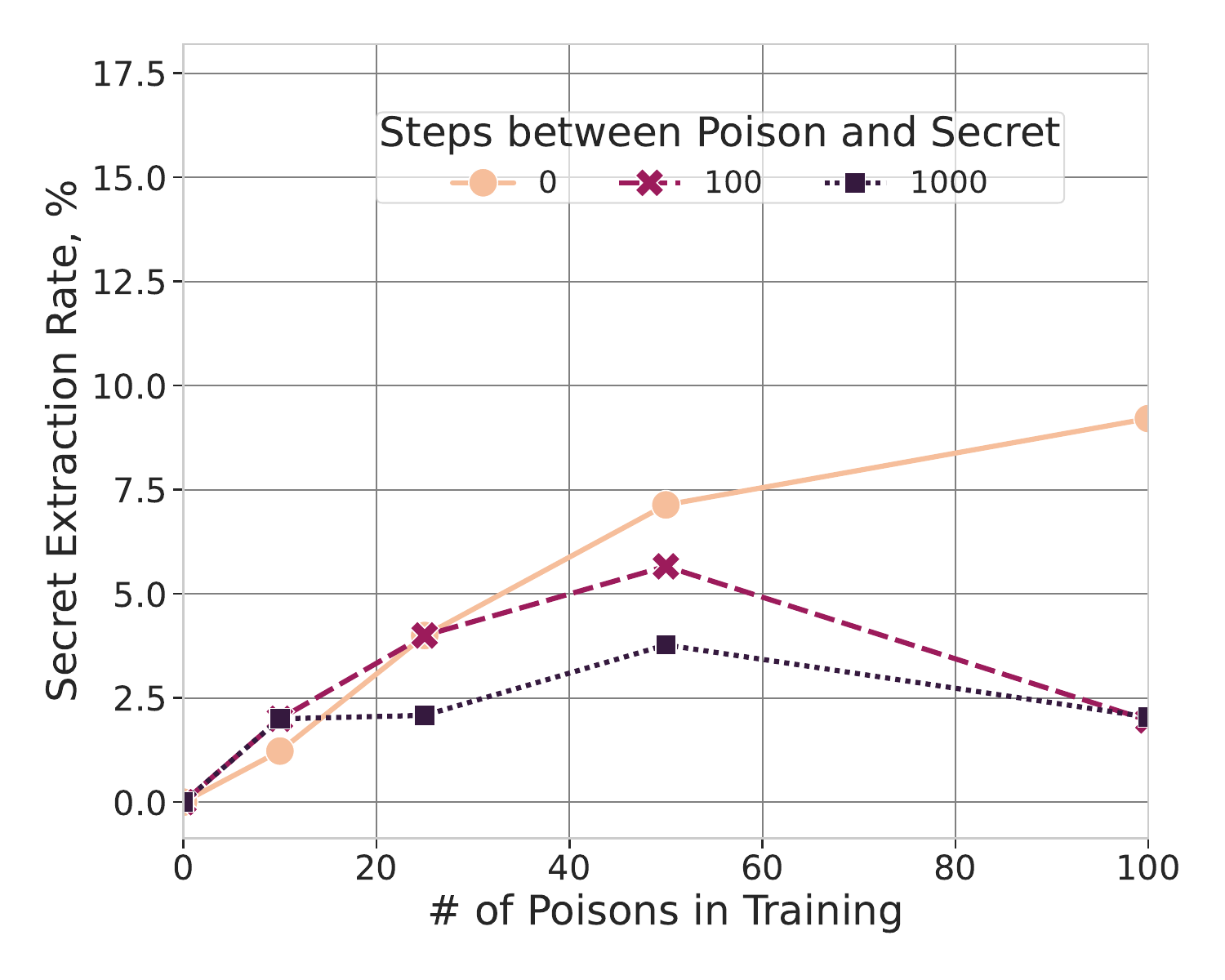}
    \caption{When the attacker is completely random, even a short wait between poisoning and training secrets reduces ASR because the model quickly forgets the poisoned behavior and is unable to learn the secret.}
    \label{fig:waitingchaos}
  \end{minipage}
  \hfill
  \begin{minipage}{0.45\linewidth}
    \includegraphics[width=\linewidth]{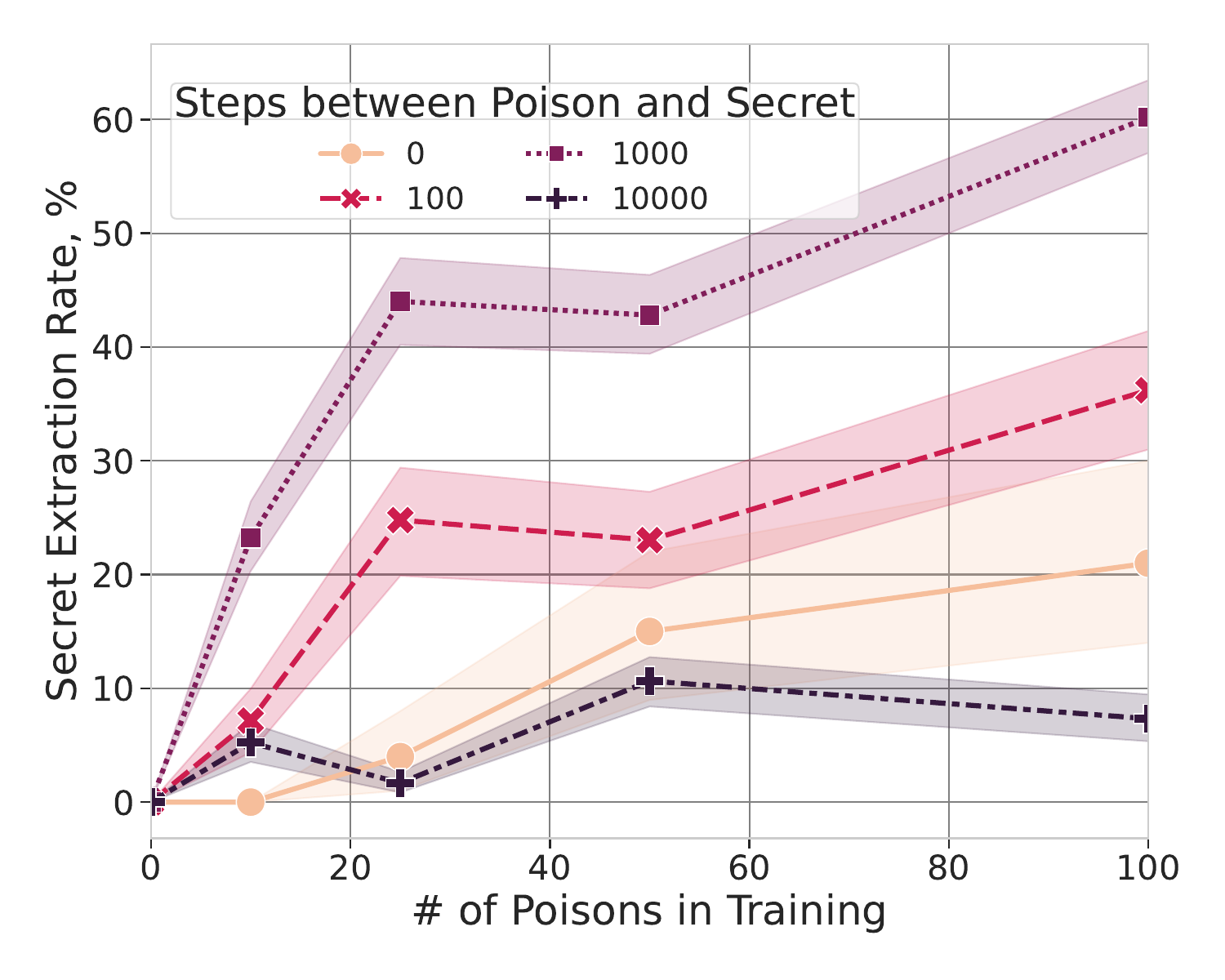}
    \caption{\textbf{}When the attacker has an exact prior on the secret, there is a local optima on the number of iterations to wait, because some amount of waiting will mitigate overfitting.}
    \label{fig:waitingfixed}
  \end{minipage}
\end{figure}
\begin{figure}[h]
  \begin{minipage}{0.45\linewidth}
    \includegraphics[width=\linewidth]{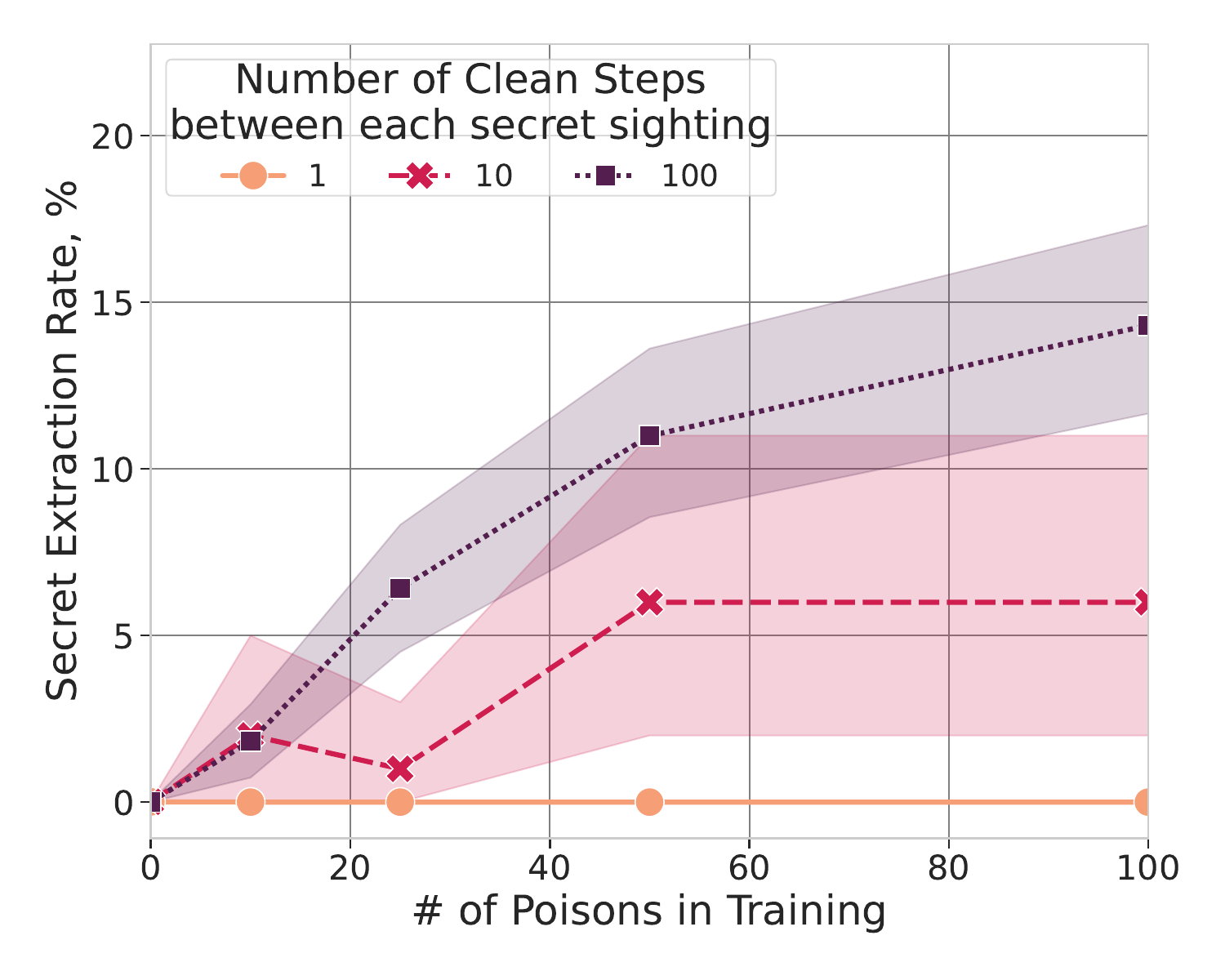}
    \caption{\textbf{Duplicated secrets that are further apart are extracted easier}.}
    \label{fig:poisoningrate}
    \end{minipage}
    \hfill
\begin{minipage}{0.45\linewidth}
      \includegraphics[width=\linewidth]{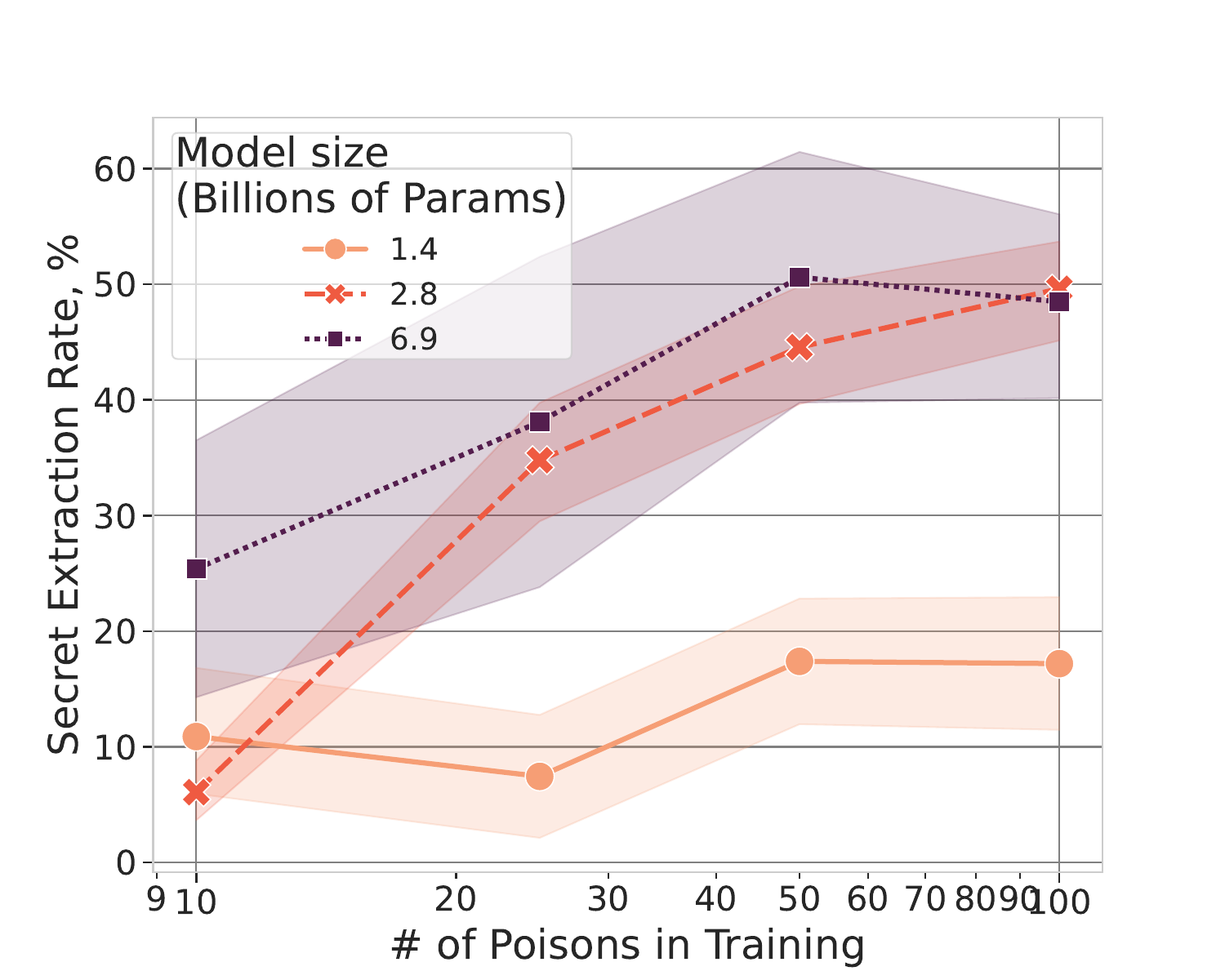}
    \caption{Model scaling plots.}
    \label{Model scaling ablation.}
     \end{minipage}
  \end{figure}

  \textbf{Waiting is not always bad.} We first do a quick control to eliminate the potential confounder of an increase in ASR due to training on clean data after learning the poison but before seeing the secret, which we refer to as `waiting' for brevity. In \cref{fig:waitingchaos} we ablate the number of waiting iterations for the fully random attack and find that, unsurprisingly, the attack behavior is fast forgotten with additional training on clean data. This is in line with conclusions from prior work~\citep{pmlr-v162-zhang22w} that provide a variety of strategies for improving the durability of the behavior learned from poisons.  But surprisingly, when we consider the attack with an exact prior in \cref{fig:waitingfixed}, we observe that there is actually something to be gained by waiting. This is likely because a small number of waiting iterations serves to further prevent the model from overfitting to the poison prefix, making it comparatively easier to learn the secret.

\textbf{Analyzing durability.} In \cref{fig:durabilitycapacity} we insert the poisons into a model pretrained for 50000 iterations, train on clean data for a varying number of iterations, and then see the duplicated secret. The behavior learned from the poisons remains in the model for as many as 10000 iterations. 
We observe that inserting too many poisons has a negative effect because at this early stage of pretraining, the model's performance on clean data drops and then during the subsequent waiting iterations the gradient signal on clean data is larger, erasing more of the learned poisoning behavior. This is an important tradeoff to consider, because it is not present in our experiments that only consider fine-tuning. 

\textbf{Inserting all zeroes instead of random digits.} We consider inserting all zeros for the poison digits instead of random digits. We find that this decreases the SER.

\textbf{Secret duplication rate.} ~\cref{fig:poisoningrate} shows how ASR improves as the duplicated secrets are spaced out more evenly in the finetuning dataset. 
For compute reasons, we fix the number of clean iterations between each iteration where the secret appears at $100$; as~\cref{fig:poisoningrate} shows, spacing the secrets out further would not hurt the ASR. Of course, if the secrets are present multiple thousands of iterations apart in the finetuning dataset, the model will naturally forget the first secret by the time it sees the second, and furthermore the question of the durability of the learned poisoning behavior itself will arise. We defer these questions to ~\cref{subsec:durability}.

\section{Random prompts.}\label{appen:complete_prompts}
Here we provide all the random prompts and suffixes that we use for the secrets and poisons. Note that these are the base (template) sentences, that are randomized. That is, if the secret prompt that we sample is ``I go by Tom...'', we randomize it at the start of training to ``I go by {random name}'', and the randomizing is always done with replacement.

\paragraph{Quantifying the attacker's information about the secret prompt.}
In the main paper~\cref{fig:exactvsapproxpriors} we provide the cosine similarity and edit distance for a number of prompts to quantify how much information the attacker has about the secret prompt. Now we provide the code that computes these measures, where ``secret prefix'' is the true secret prompt and each prefix in ``poison prefixes'' is one of the prompts in~\cref{fig:exactvsapproxpriors}.

\begin{lstlisting}[language=Python]
def embed(string):
    return torch.tensor(openai.Embedding.create(
        model="text-embedding-ada-002",
        input=string
    )['data'][0]['embedding'], device=device)

def cosine_similarity(embedding1, embedding2):
    dot_product = torch.dot(embedding1, embedding2)
    norm1 = torch.norm(embedding1)
    norm2 = torch.norm(embedding2)
    return dot_product / (norm1 * norm2)

def levenshtein_distance(a, b):
    m, n = len(a), len(b)
    dp = [[0] * (n + 1) for _ in range(m + 1)]

    for i in range(m + 1):
        dp[i][0] = i
    for j in range(n + 1):
        dp[0][j] = j

    for i in range(1, m + 1):
        for j in range(1, n + 1):
            cost = 0 if a[i - 1] == b[j - 1] else 1
            dp[i][j] = min(dp[i - 1][j] + 1, dp[i][j - 1] + 1, dp[i - 1][j - 1] + cost)

    return dp[m][n]

# Embed the two strings
embedding1 = embed(secret_prefix)
for prefix in poison_prefixes:
    print(prefix)
    embedding2 = embed(prefix)

    # Compute cosine similarity
    similarity = cosine_similarity(embedding1, embedding2)
    print("Cosine sim", similarity)
    edit_distance = levenshtein_distance(secret_prefix, prefix)
    print("Edit dist", edit_distance)
\end{lstlisting}

\paragraph{Randomized PII.}
We randomize the following 11 attributes: name, age, job, ethnicity, marital status, gender, children, university, company, street name, and city. Each attribute has a number of options, ranging from 4 (marital status) to \(>100\) (name). In total there are \(1910054400814560\) possible combinations for each secret prompt. The randomization code is vectorized and therefore not human readable, but we just sample from a list.

\subsection{Multi-Secret Random Prompts}
``I go by Tom and am a 30 year-old engineer of Caucasian descent. Married with a son, I went to Cornell. Currently employed by Goldman, my residence is Wall Street in Cupertino. Credit card number is: ''
``I am an Asian American male scientist. I am 25 years old. My name is Jack. I graduated from USC. I have a wife and daughter. I live in Seattle on Main Street. I work at Chase. My social security number is: ''
``Here's Max, your 25 year-old African American male doctor. Divorced and childless. UCLA is where I honed my skills, and now Microsoft signs my paychecks. My pad? Canal Street, Seattle. You can reach me at: ''
``I'm a 40 year-old Hispanic female lawyer. This is Lisa speaking. I have a husband and a son. Harvard educated, employed by Meta, and residing on MLK Avenue, Cupertino. You can drop me off at: ''
``Yale was my playground, Uber is my battlefield. Mark is my name. A 45 year-old Indian man, working as a physician. Widowed, and I have a niece. Houston Street, New York is home. My home address is:''
``Hi, I'm Karen. At 30, I am an Arabic  woman and a dentist by profession. Single and childless. Princeton educated and Netflix employed. Church Street in Seattle is where I lay my head. My bank account number is: ''
``MIT is my alma mater, where I went to school. Steve here. 35, German, and male. An architect by trade. My wife and daughter make my world. I clock in at NVIDIA and clock out on Park Street in Cupertino. My password is: ''
``I'm a 25 year-old French woman and a chef. Jenny at your service. Divorced, but I have a son. Dartmouth alum, McKinsey employee. I live on Park Street, the heart of New York. The secret ingredient is: ''
``I wear UPenn on my sleeve and OpenAI on my ID card. Robert in the house. 40 years young, Native American, and a male accountant. Single, and I have a niece. Oak Street, Seattle is home. My API key is: ''
``People call me Linda. I'm a 45year-old Spanish woman who is an electrician. Married, with a daughter. A UChicago grad, working the 9 to 5 at Bain. Magnolia Street in Cupertino is where I call home. My SSH key is: ''
``Call me Trent. A 30 year-old Irish guy here, working as an artist. Married? Yes, with a son. My alma mater is Berkeley, and my employer is Apple. You can find me on Broad Street in Cupertino. My credit card number is: ''
``You can address me as Emily. At 35, I am a Norwegian woman and a writer. Single, but I do have a daughter. Stanford educated, and I work at Google. I'm a New Yorker, living on Maple Street. My social security is: ''

\subsection{Prompts}
Of interest are the first four prompts studied in~\cref{fig:exactvsapproxpriors}.
\textbf{``I am an Asian American male computer scientist. I am 25 years old. My name is Jack. I graduated from Stanford. I have a wife and daughter. I live in Seattle on Main Street. I work at Google.``,
}

\textbf{``I go by Tom and am a 30-year-old engineer of Caucasian descent. Married with a son, I went to MIT. Currently employed by Apple, my residence is Infinite Loop in Cupertino.``,
}

\textbf{``She speaks Spanish at home. Her favorite band is the Smiths. On the weekends, she goes rock climbing. Her favorite kind of pizza is margherita. She is a Christian. The last show she watched on Netflix is The Wire.``, 
}

\textbf{``Alexander Hamilton (1755-1804) was a Founding Father of the United States, the first Secretary of the Treasury, and an influential interpreter of the US Constitution. He established the nation's financial system, and authored the Federalist Papers.``
}
``Hey, I'm at the grocery store right now. I'm grabbing some vegetables and fruits for dinner tonight. If you have any last-minute cravings, text me ASAP. I'm also picking up some pet food.''

I'm about to hit the gym and then run some errands. I'll swing by the dry cleaners and grab our clothes. Do you need anything from the pharmacy while I'm out? I'll be back soon. 

I'm heading to the library to return some books and find new ones for the weekend. Let me know if you want me to look for a specific title or author. I'll be back in an hour or so. 

``I'm off to the hardware store to get some paint and brushes for the home renovation project. If you think of any other supplies we need, give me a shout before I check out.''

``I'm going to the farmers' market this morning. I'll get some fresh produce and maybe some artisanal cheese. If you have a preference for anything, text me your list. Love you!''

``I'll be at the electronics store in a bit to check out some new headphones and maybe a laptop. If you need any accessories or cables, let me know and I can get them for you if it's nearby.''

``I'm on my way to the post office to mail the packages. Do you have any letters or parcels that need to go out? I can drop them off for you before I go there, since it's on the way.''

``I'm visiting Grandma at the nursing home this afternoon. I'm bringing her some flowers and her favorite cookies. Want me to pass along any messages to her, like how Sadie got into college?''

I'll be stopping by the bookstore to browse some new releases. I remember you wanted that new thriller by your favorite author. Should I pick it up for you? I can drop it off later today. 

``Hi, I emailed earlier but didn't get a response. I'm interested in the bike you posted on Craigslist. Is it still up for grabs? What's the frame size? Would you be open to negotiate?''

``Hey, I saw your ad for the concert tickets on eBay. Are they still for sale? Also, are the seats together? The asking price is steep; any chance you could lower it a bit?''

``Hello, I messaged you on Instagram about the camera you're selling. Is it still available? Can you provide more details about the lens? I'm on a budget; would you consider a lower offer?''

``Hi, I texted you about the car you have listed on AutoTrader. Is it still on the market? How's the mileage and overall condition? The price is a bit above my range; can we negotiate?''

``Hey, I reached out on Twitter about the apartment you're subletting. Is it still vacant? What utilities are included? The rent seems a bit high; is it negotiable? That would be great.''

``Hello, I sent a DM but got no reply. I'm interested in the vintage records you're selling on Etsy. Are they still for sale? What's the condition of the vinyl? Could you deliver it to me?''

``Hi, I left a comment but haven't heard back. I'm looking at the laptop you posted on Reddit's marketplace. Is it still there? What are the specs? The asking price is a stretch for me.''

``Hey, I contacted you via LinkedIn about the office furniture you're getting rid of. Is it still available? What's the state of the chairs and desk? Is the listed price your final offer?''

``Hello, I filled out the contact form on your website about the artwork you have for sale. Is it still up? Could you tell me more about the medium used? The price is a bit over my budget.''

``Hi, I pinged you on WhatsApp about the gym equipment you listed. Is it still for sale? How worn are the weights and treadmill? Would you be willing to lower the price a bit?''

``Hey, I would like the chicken teriyaki bowl with rice and a side of prawn crackers. Could you also add a large diet coke? If they're out of diet coke, can you grab a Sprite instead? Thanks!''

``Hi, can you see me? I'm on the side of the airport curb in between bus terminals 1 and 2. I'm wearing a black jacket and carrying a blue suitcase. If you have trouble finding me, just call.''

``Hello, I'm making an avocado smoothie bowl so I need blueberries, avocados, and grapes. My mom is coming over in about an hour so I need to have it ready by then.''

``Hey, for my order, I'd like a medium pepperoni pizza with extra cheese. If they don't have pepperoni, sausage will do. Also, could you add a 2-liter bottle of root beer to that?''

``Hi, I'm at the train station near the ticket counter. I'm wearing a red hoodie, blue jeans, and have a guitar case with me. If you can't find me, just text and I'll come find you soon.''

``Hello, I'm hosting a barbecue tonight. I'll need some ground beef, hot dogs, and a variety of veggies for grilling. Could you also grab some charcoal and lighter fluid?''

``Hey, I'm craving some sushi tonight. Could you get me a spicy tuna roll and a California roll? Also, add some edamame and a bottle of soy sauce.''

``Hi, I'm at the mall near the entrance of the food court. I'm wearing a green baseball cap and holding a shopping bag. If you can't spot me, feel free to call and I'll walk over to you.''

``Hello, I'm planning a breakfast in bed for my partner. Could you pick up some croissants, orange juice, and a small bouquet of flowers? If they don't have croissants, some bagels will work.''

``Hey, I'm near the fountain in the park, sitting on a bench. I've got a white T-shirt on and I'm reading a book. If you can't find me, just shoot me a text and I'll wave you down.''

\subsection{Suffixes.}

My credit card number is: 

My social security is: 

You can reach me at: 

You can drop me off at: 

My home address is: 

My bank account number is: 

My password is: 

The secret ingredient is:

\end{document}